\documentclass[fleqn,usenatbib]{mnras}

\usepackage[T1]{fontenc}
\usepackage{ae,aecompl}
\usepackage{xtab}
\usepackage{courier}
\usepackage{multirow}
\usepackage{times}
\usepackage{upgreek}
\usepackage{graphicx}
\usepackage{epstopdf}
\usepackage{amssymb}
\usepackage{amsmath}
\usepackage[section]{placeins}

\usepackage[T1]{fontenc}
\usepackage{ae,aecompl}

\usepackage{dirtytalk}
\usepackage{comment}
\newcommand{\tess}{{\em TESS}}
 
 \newcommand{\cmss}{\,\mbox{$\mbox{cm}\,\mbox{s}^{-2}$}}    % centimetres/second^2

\title[TESS sdBVs in the southern ecliptic hemisphere]{A search for variable subdwarf B stars in TESS Full Frame Images\\
I. Variable objects in the southern ecliptic hemisphere}
\author[S.K.\,Sahoo et al.]{
S.K.\,Sahoo,$^{1,2}$\thanks{E-mail: sumanta.kumar27@gmail.com}
A.S.\,Baran,$^{1,3}$
S.\,Sanjayan$^{1,2}$
and J.\,Ostrowski$^{1}$
\\
$^{1}$ARDASTELLA Research Group, Institute of Physics, Pedagogical University of Krakow, ul. Podchor\c{a}\.zych 2, 30-084 Krak\'ow, Poland\\
$^{2}$Nicolaus Copernicus Astronomical Centre of the Polish Academy of Sciences, ul. Bartycka 18, 00-716 Warsaw, Poland\\
$^{3}$Department of Physics, Astronomy, and Materials Science, Missouri State University, Springfield, MO 65897, USA\\
}

\date{Accepted XXX. Received YYY; in original form ZZZ}

\pubyear{2020}

\begin{document}
\label{firstpage}
\pagerange{\pageref{firstpage}--\pageref{lastpage}}
\maketitle

% Abstract of the paper
\begin{abstract}
We report the results of our search for pulsating subdwarf B stars in Full Frame Images, sampled at 30\,min cadence and collected during Year 1 of the {\it TESS} mission. Year 1 covers most of the southern ecliptic hemisphere. The sample of objects we checked for pulsations was selected from a subdwarf B stars database available to public. Only two positive detections have been achieved, however, as a by-product of our search we found 1807 variable objects, most of them not classified, hence their specific variability class cannot be confirmed at this stage. Our preliminary discoveries include: two new subdwarf B (sdB) pulsators, 26 variables with known sdB spectra, 83 non-classified pulsating stars, 83 eclipsing binaries (detached and semi-detached), a mix of 1535 pulsators and non-eclipsing binaries, two novae, and 77 variables with known (non-sdB) spectral classification. Among eclipsing binaries we identified two known HW\,Vir systems and four new candidates. The amplitude spectra of the two sdB pulsators are not rich in modes, but we derive estimates of the modal degree for one of them. In addition, we selected five sdBV candidates for mode identification among 83 pulsators and describe our results based on this preliminary analysis. Further progress will require spectral classification of the newly discovered variable stars, which hopefully include more subdwarf B stars.
\end{abstract}

% Select between one and six entries from the list of approved keywords.
% Don't make up new ones.
\begin{keywords}
stars: binaries: eclipsing -- stars: oscillations (including pulsations) -- stars: subdwarfs -- stars: variables: general
\end{keywords}

%%%%%%%%%%%%%%%%%%%%%%%%%%%%%%%%%%%%%%%%%%%%%%%%%%

%%%%%%%%%%%%%%%%% BODY OF PAPER %%%%%%%%%%%%%%%%%%

\section{Introduction}
Subdwarf B\,(sdB) stars are identified as objects located at the blue end of the horizontal branch in the Hertzsprung-Russell diagram \citep{heber16}. These stars are compact in size, with surface gravities, $\log(g/\cmss)$, of 5.0 to 5.8, which translates into radii of 0.15\,--\,0.35\,R$_{\sun}$. SdBs are blue due to their high effective surface temperature\,(T$_{\rm eff}$) ranging between 20,000 and 40,000\,K. Such a high temperature makes them candidates for the ionizing sources of interstellar gas at high galactic latitudes \citep{deboer85} and mostly responsible for the ultraviolet upturn phenomenon in early-type galaxies \citep{brown97}. The sdB stars have masses 0.47\,M$_{\sun}$ on average, which is sometimes called the canonical mass \citep{heber16}.

After the discovery of pulsating sdBs (sdBV) observationally by \citet{kilkenny97} and theoretically by \citet{charpinet97}, asteroseismology became the major tool to investigate the interior of sdBs. The pulsations were found at both low and high frequencies. The low frequencies (long periods of hours) are explained by gravity modes, while the high frequencies (short periods of minutes) are explained by pressure modes \citet{fontaine03}. Since these stars were discovered only recently, it was essential to make an effort in more discoveries to increase the sample of sdBVs to a statistically significant number. First discoveries were made from the ground and detections were limited to sdBVs showing pressure modes. They were easier to detect because of their higher observable amplitudes and shorter periods. sdBVs with gravity modes were also detected from the ground, but the actual number of these discoveries was never published. Only a handful sdBVs were found to be pulsating in both types of modes, with Balloon\,090100001 being the best example \citep{baran05}. Overall, there are about 50 sdBVs found only from the ground. Thanks to the {\it Kepler} spacecraft \citep{borucki10} we found more sdBVs. Most of them pulsate in gravity modes and many of them show both types of pulsations, which means that hybrid behaviour is common among sdBVs. To date, there are around 130 sdBV found in both ground and space data \citep{holdsworth17,reed18b}. SdBVs are found in both open \citep{reed12} and globular \citep{randall09} clusters but mostly as Galactic-field counterparts. While ground-based discoveries were made all over the sky, {\it Kepler} discoveries were limited, first to the fixed region during the original mission, and second to the ecliptic plane during {\it K2} mission. We lack a complete all-sky search for sdBVs. It prevented us from a comprehensive analysis of pulsation properties correlated with stellar population to conclude the location of instability strip(s) or physical parameters (especially masses).

At present the {\it Kepler} successor, the Transiting Exoplanet Survey Satellite \citep[TESS,][]{ricker14} is an all-sky survey whose primary goal is to detect exoplanets orbiting around nearby bright stars using the transit method. As a by-product, time-series photometry of nearly 20,000 additional targets for astrophysical research are produced (pre-defined targets). These data are taken either in the long cadence (LC) or short cadence (SC) modes. The LC is 30\,min, while the SC is 2\,min. What makes {\it TESS} different from {\it Kepler} is that full-frame images (FFI) taken with LC are all downloaded and available to public. This is a significant source of time-series data, since almost the entire sky is sampled at the 30\,min resolution. Since only for pre-defined targets, time-series data are prepared by the in-house pipeline, the other objects detected in the FFIs need special data processing. Handling all objects detectable in the FFIs, although desirable, is technically difficult (processing would take months if not years), one can focus on a selected target type only, limiting the time needed for data processing and time-series delivery. During Year 1 and 2, {\it TESS} covered nearly 85\% of the sky. The entire {\it TESS} field is divided into 26 sectors and each sector is monitored for 27\,days. After first 13 sectors in the southern ecliptic hemisphere were completed, {\it TESS} switched to the northern ecliptic hemisphere.

The main goal of our work was to select sdBs and sdB candidates that were not included in the pre-defined target list, produce time-series data directly from FFIs and make mode identifications of the best suitable cases showing pulsations. An all-sky search for sdBVs, along with {\it Gaia} parallaxes will tell us about the distribution of these stars in the Galaxy and contribute toward pulsation-stellar population relationship. In Section\,\ref{gaia} we describe the source of our targets, the selection process and data processing. In Section\,\ref{results} we present objects with found variables. Section\,\ref{modeid} reports our mode identification effort, followed by Section\,\ref{summary} that summarizes our results.

\section{Target selection and flux extraction}
\label{gaia}
We used the sdB database described by \citet{geier20} that we consider the most updated database of confirmed sdB as well as sdB candidates. It was prepared based on {\it Gaia} mission \citep{gaia18a}, specifically ESA {\it Gaia} Data Release 2 (DR2) and several ground-based multi-band photometry surveys. Color indices, absolute magnitudes and reduced proper motions were used to select the most suitable candidates. The database is limited to {\it Gaia} G mag\,=\,19 and contains 39\,800 objects. From this sample we selected targets located in the southern ecliptic hemisphere and covered by {\it TESS} silicons. Using {\it Gaia} IDs and target coordinates we applied TOPCAT \citep{taylor05} to reject targets that are assigned to be observed in the SC mode. Then, we used {\it Tesscut} \citep{brasseur19} to collect sector information targets in our sample will be observed in, and targets with no sector assignment were also rejected. Finally, we filtered targets assigned to sector 1-13 only, and we ended up with 21,879 targets. It turned out that 1237 targets have no useful data, so these were also rejected from our sample, resulting in 20,642 targets. For completeness, we have included targets with non-sdB spectral classification. It may also happen that some of these already classified objects will be reclassified as sdBs with further analysis. In addition, other researchers may find it useful to have these objects identified as variables.

%Before proceeding, we divided this list in terms of spectral type mentioned in Geier's database. First sublist contains the targets with mentioned spectral type \say{sdB}. This sublist targets have the highest probability to be sdBVs. We found total 532 targets in this category. The second sublist contains targets with mentioned spectral type containing phrase \say{sd} and excluding \say{sdB}. These targets have also higher probability to be sdBVs. There are 211 targets in this category. The third sublist contains rest of the targets which have some other spectral type mentioned in the list or there is no spectral classification. There are 21\,136 targets in this category.

We used the {\it Eleanor} \citep{feinstein19} that is an open source python framework developed for downloading, analysis, and visualization of data directly from the {\it TESS} FFIs. It is able to extract corrected time-series data for a given object. As input it takes the TIC ID, {\it Gaia} ID, or coordinates along with observed {\it TESS} sector information, and returns with a table FITS file with time-series data, optimal aperture shape applied.

The {\it Eleanor} works in two steps. First, it detects a target and creates a target pixel file (TPF), then it does aperture photometry to extract the flux and create time-series data. {\it Eleanor} works directly with the "Barbara A. Mikulski Archive for Space Telescopes" (MAST) to download all necessary data for a given target. We specified a square target mask of 15\,pixels on side and a square background mask of 31\,pixels on side. It delivers raw and corrected time-series data. The raw data is a simple sky-subtracted simple aperture photometry, which is basically a sum of all flux within an optimal aperture for each timestamp. The corrected data account for known satellite artifacts. We extracted both data sets and have chosen the one that shows better signal to noise ratio. %In general, we selected the corrected data, however in several cases the flux variations are so large that the correction process removed part of the real signal. In those cases, we selected the raw data (with the artifacts not removed).
All details on the {\it Eleanor} can be found in the reference paper by \citet{feinstein19}.

We used {\it lightkurve} python package \citep{hedges18} to detrend and remove outliers from both raw and corrected time-series data. We clipped the data at 4$\sigma$ and de-trended long term variations (longer than two days). Then, we cleaned these data again by removing data points that were not clipped, but still significantly deviated from the expected trend. These were data points at the beginning and/or end of each satellite orbit that are caused by unstable satellite temperature. Finally, we normalized fluxes by calculating $(f/\bar{f}-1)*1000$ reporting amplitudes in {\it parts per thousand} (ppt). We stress that the {\it Eleanor} does not remove contributions from neighboring objects, which results in overestimated average flux and diluted amplitude of flux variations. Therefore, the amplitudes are not realistic and are affected by our case-to-case definition of a flux zero point accurately. In case of sdBV stars, the amplitudes are not important, but accurate modelling of eclipsing binaries or classical pulsators may require further individual effort to pull out the fluxes with preserved amplitudes.

We calculated amplitude spectra for each target and we used them to search for flux variations. These spectra help detect variations even if the time-series data show no significant (by eye) variations. We accept a positive detection only if peaks in the spectra meet our signal-to-noise (S/N) ratio of 4.5 \citep{baran15}. Since we work with the long cadence data, the Nyquist frequency equals 277\,${\mu}$Hz (24c/d), which translates to 30\,min period. This limits our search to gravity mode sdBVs only.

\section{Results of our flux variation search}
\label{results}
We detected significant flux variations in 1807 objects, out of which 28 are classified as sdBs, 2 as subdwarfs (sd), 77 as non-sdBs, and 1619 are not spectroscopically classified. To identify sdB objects we used the sdB database \citet{geier20}, the LAMOST spectroscopic survey \citep{lei18,lei19,lei19a,luo19}, the Evryscope survey \citep{Ratzloff20} and the Simbad database \citep{wenger2000}. The large square pixels, 21\,arcsec on side, cause serious issues in crowded regions of the sky, since an optimal aperture may contain neighboring objects. We over-plotted optimal apertures of all our objects, adopted by the {\it Eleanor}, on top of the Digitized Sky Survey (DSS) color images using the {\it Aladin} \citep{bonnarel2000}. If the aperture overlaps with additional sources, and we are not positive about which object a flux variation comes from, we made remarks in Tables\,\ref{tab:table1}--\ref{tab:table7} that provide basic information on our target findings and possible contamination. In case an optimal aperture region contains more than five objects we marked it with \say{crowded region} remark, while in case the optimal aperture is densely covered by stars, {\it e.g.} a cluster region, we marked it with \say{very crowded region}. If a few objects\,(<\,5) were spotted, we specified the number of objects within a radius or in case of a single object within optimal aperture, we provided the distance to the object. If possible, we provided designations of such objects. In addition, some of the objects may have already been discovered and published. If we found that a given target is already known as a variable star, we provided a relevant reference. We included these objects for completeness and to report on the frequency of variable stars found in the {\it TESS} data.

\subsection{SdB stars}
\begin{figure*}
\includegraphics[width=0.85\textwidth]{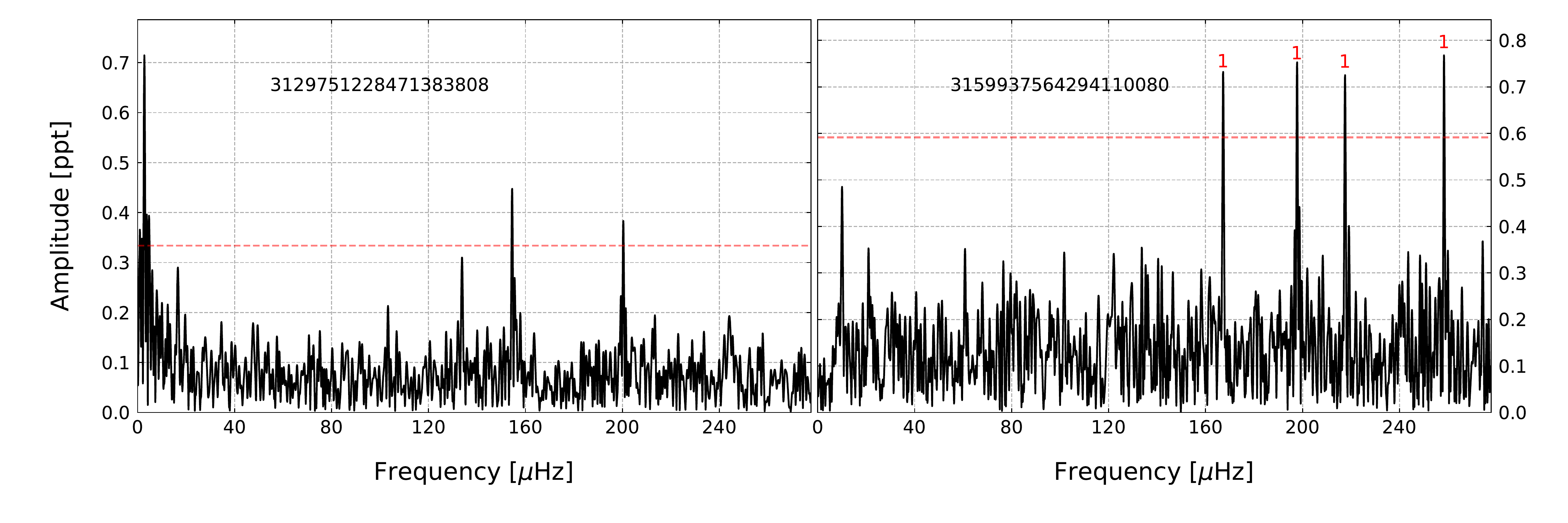}
\caption{Amplitude spectra of two sdBV stars. Horizontal red dashed line indicates $4.5\,\sigma$ detection threshold (all relevant figures). In the right panel we added our guesses on the modal degree based on the period spacing.}
\label{fig:sdBV}
\end{figure*}

We found 28 variable objects classified as \say{sdB} and we list them in Table\,\ref{tab:table1}. Two objects, {\it Gaia}\,DR2\,3129751228471383808 and 3159937564294110080, show multiple peaks in their amplitude spectra (Figure\,\ref{fig:sdBV}), which we interpret as g-modes. Both objects were first observed by \citet{hog2000} and the spectral classifications were made by \citet{luo19} and \citet{lei18}.

\begin{figure*}
\includegraphics[width=0.85\textwidth]{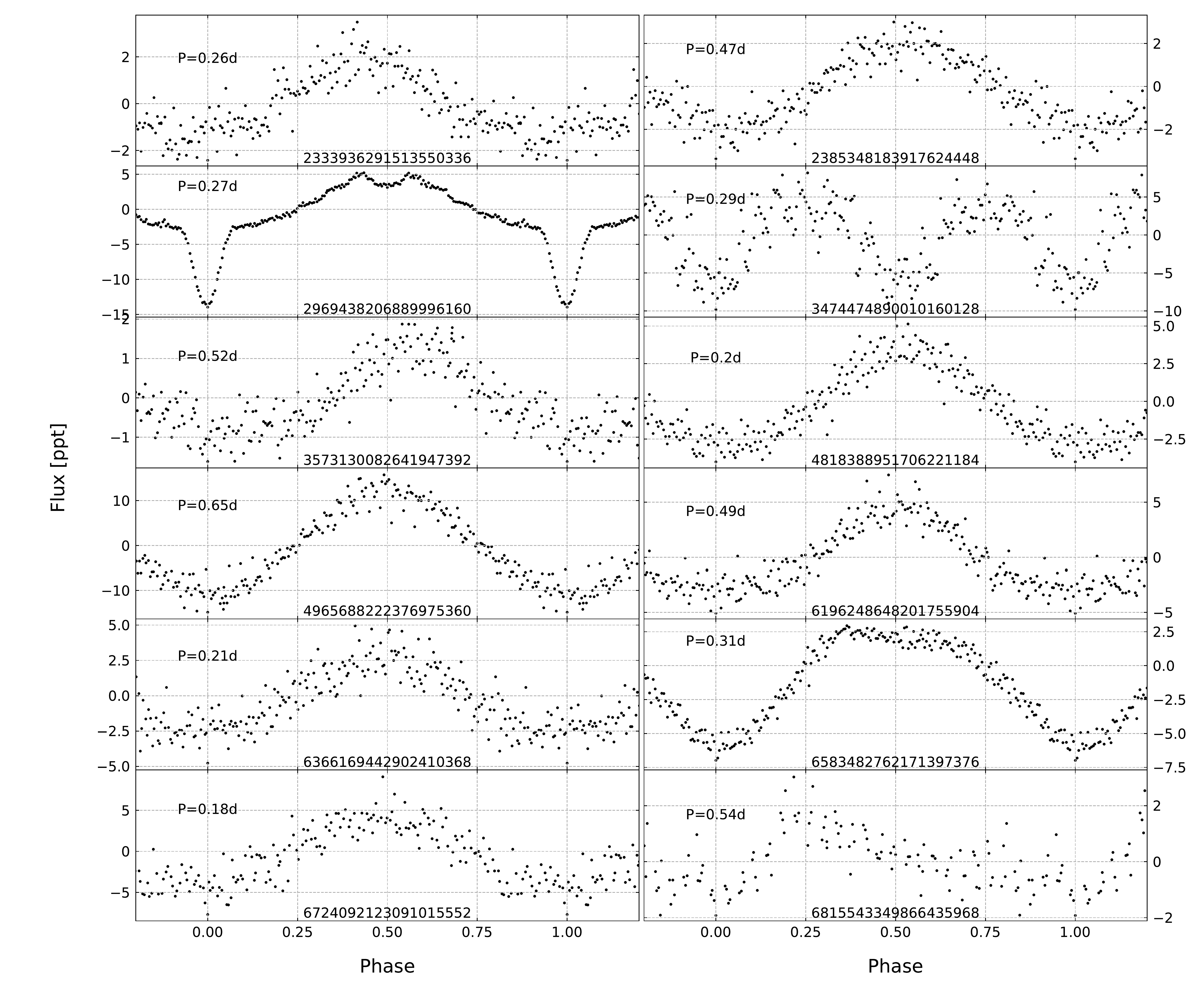}
\caption{Phased and binned time-series data of a sample of sdBs that show relatively large amplitude flux variations.}
\label{fig:sdBp}
\end{figure*}

The other 26 objects show light variation typical of binaries. All objects (here and later on) that show mono-periodic, most likely binary, behavior with large amplitude flux variations were phased and the orbital periods were cited in the relevant figures and tables. If a flux variation is easily seen in the light curves we phased and binned data and these are plotted in Figure\,\ref{fig:sdBp}. This figure contains candidates for reflection binaries (a sdB and a low mass main sequence companion) and ellipsoidal binaries (a sdB and a white dwarf). One of the eclipsing binaries, {\it Gaia}\,DR2\,2969438206889996160, which has just recently been published by \citet{Ratzloff20}, is an HW\,Vir system \citep{wood93,baran18}.

\begin{figure*}
\includegraphics[width=0.85\textwidth]{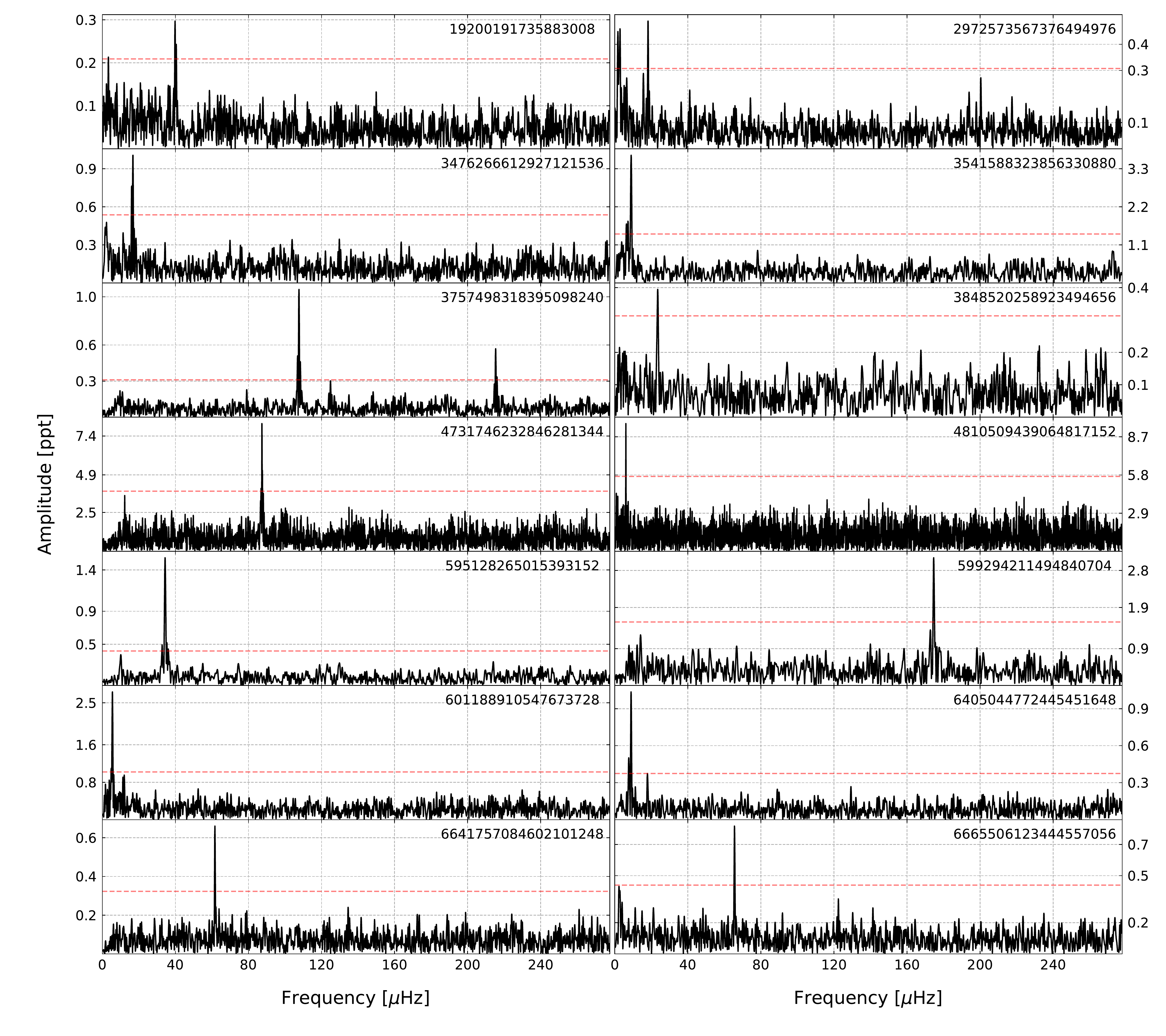}
\caption{Amplitude spectra of a sample of sdBs that show low amplitude flux variations.}
\label{fig:sdBf}
\end{figure*}

Figure\,\ref{fig:sdBf} includes 14 variable sdB stars that show low amplitude variation that are detectable in amplitude spectra. Even though the peaks we detected are in the g-mode region, one or two peaks can be interpreted as either pulsations or binarity. The only exception could be {\it Gaia}\,DR2\,599294211494840704 that shows a single peak between 160 and 200\,$\upmu$Hz. {\it Gaia}\,DR2\,3757498318395098240 shows two peaks above 80\,$\upmu$Hz, but the higher frequency peak is a harmonic, hence we claim it to be a signature of binarity, as well.

\begin{table*}
	\centering
	\caption{Basic information of 28 objects classified as sdBs. Time-series data or amplitude spectra of these objects are plotted in Figs.\,\ref{fig:sdBV},\ref{fig:sdBp} and \ref{fig:sdBf}. Objects in all tables and figures are sorted by {\it Gaia} ID.}
	\label{tab:table1}
	\resizebox{0.85\textwidth}{!}{\begin{tabular}{lllllll}
		\hline
		{\it Gaia} DR2 & TIC & Name & G mag & Sector & Period[d] & Remarks\\
		\hline
19200191735883008 & 337216760 & PG\,0229+064 & 11.9242 & 4 & 0.2906 & -\\
2333936291513550336 & 12379252 & Ton\,S138 & 16.0148 & 2 & 0.2648 & -\\
2385348183917624448 & 9035375 & PHL\,460 & 12.2071 & 2 & 0.4734 & object 1.3' away\\
2969438206889996160 & 139397815 & HW\,Vir type & 13.6079 & 5-6 & 0.2746 & \citet{Ratzloff20}\\
2972573567376494976 & 146340999 & EC\,04542-2320 & 16.1603 & 5 & 0.6377 & -\\
3129751228471383808 & 237597052 & TYC\,161-49-1 & 11.1375 & 6 & 0.05-0.1 & sdBV candidate\\
3159937564294110080 & 262753627 & TYC\,770-941-1 & 12.4615 & 7 & 0.04-0.08 & sdBV candidate\\
3474474890010160128 & 402107174 & EC\,12067-2747 & 15.8566 & 10 & 0.2923 & -\\
3476266612927121536 & 443619867 & HE\,1221-2618 & 14.6445 & 10 & 0.6925 & -\\
3541588323856330880 & 219512715 & EC\,11362-2049 & 14.1469 & 9 & 1.2933 & -\\
3573130082641947392 & 386644511 & PG\,1145-135 & 14.2710 & 9 & 0.5239 & -\\
3757498318395098240 & 902644573 & PG\,1039-119 & 16.4582 & 9 & 0.1075 & -\\
3848520258923494656 & 275358553 & PG\,0957+037 & 15.4126 & 8 & 0.4928 & -\\
4731746232846281344 & 198005084 & EC\,03572-5455 & 16.2585 & 2-4 & 0.1325 & -\\
4810509439064817152 & 200323355 & EC\,05043-4538 & 16.4590 & 4-6 & 1.9133 & -\\
4818388951706221184 & 77372867 & 2MASS\,J04512188-3743059 & 16.0941 & 4-5 & 0.2007 & -\\
4965688222376975360 & 49593787 & - & 13.4408 & 3 & 0.6467 & \citet{vos18}\\
595128265015393152 & 366656123 & 2MASS\,J08412266+0630294 & 14.8264 & 7 & 0.337 & -\\
599294211494840704 & 366353515 & PTF\,1J082340.04+081936.5 & 14.7016 & 7 & 0.0663 & \citet{kupfer17}\\
601188910547673728 & 800381314 & 2MASS\,J08251803+1131062 & 14.6493 & 7 & 2.0941 & \citet{Boudreaux17}\\
6196248648201755904 & 6116091 & HE\,1318-2111 & 14.7001 & 10 & 0.4879 & \citet{kupfer15}\\
6366169442902410368 & 265124418 & JL\,24 & 15.2841 & 13 & 0.2091 & -\\
6405044772445451648 & 234287962 & EC\,22209-6344 & 15.3326 & 1 & 1.3015 & -\\
6583482762171397376 & 159735013 & CD-3914181 & 10.9537 & 1 & 0.3054 & CMC\,316837 1.3" away\\
6641757084602101248 & 320055780 & EC\,19301-5523 & 15.9155 & 13 & 0.1878 & -\\
6665506123444557056 & 1990074078 & EC\,20043-5310 & 15.2382 & 13 & 0.1767 & object 2.6" away\\
6724092123091015552 & 86141703 & - & 13.4682 & 13 & 0.1782 & \citet{Ratzloff20}\\
6815543349866435968 & 302114308 & EC\,21271-2412 & 15.9811 & 1 & 0.5418 & -\\
        \hline
	\end{tabular}}
\end{table*}

\subsection{Pulsators}
\label{pulsators}
\begin{figure*}
\includegraphics[width=0.85\textwidth]{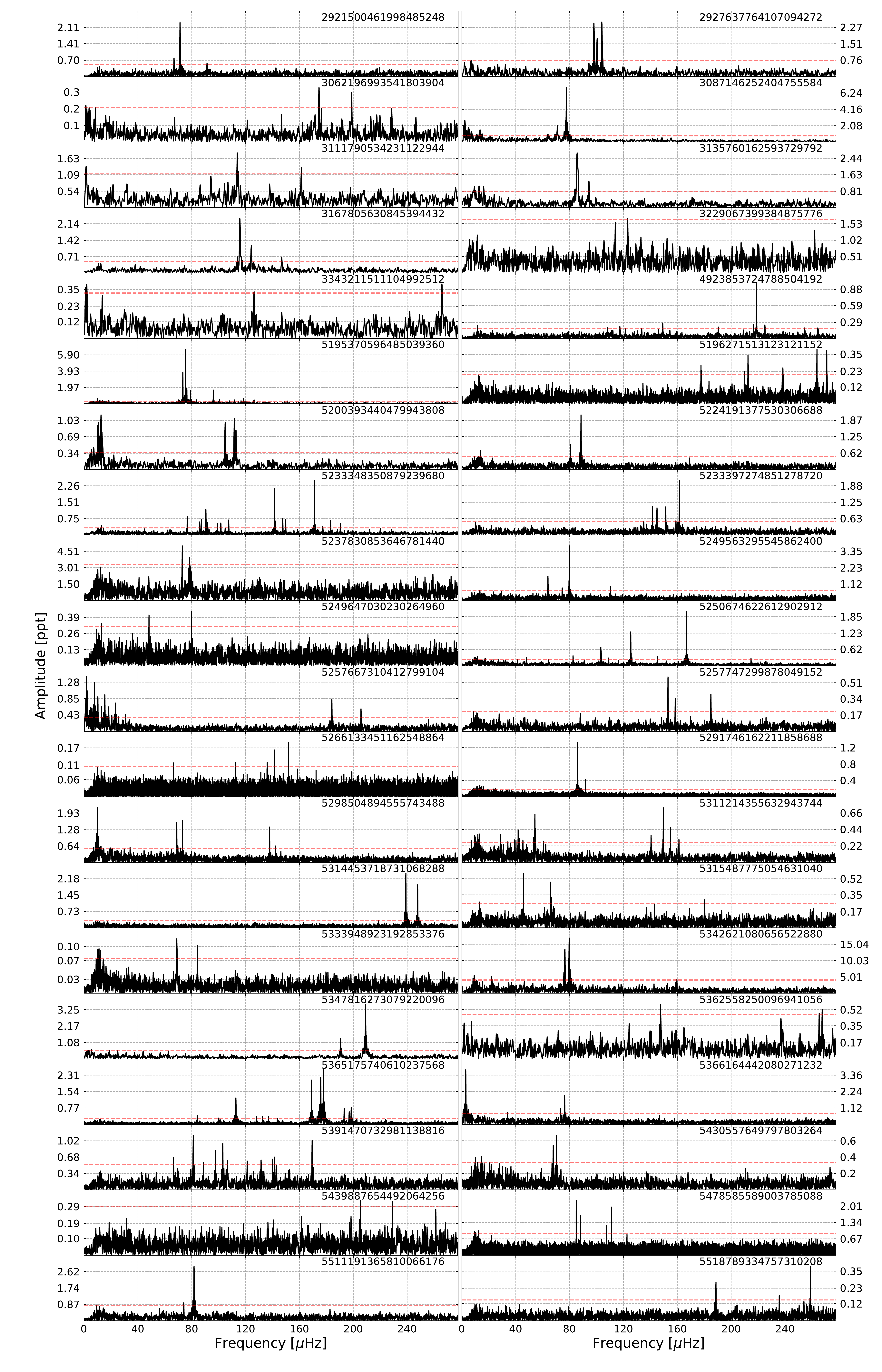}
\caption{Amplitude spectra of pulsator candidates that are not spectroscopically classified.}
\label{fig:pulsators1}
\end{figure*}

\begin{figure*}
\includegraphics[width=0.85\textwidth]{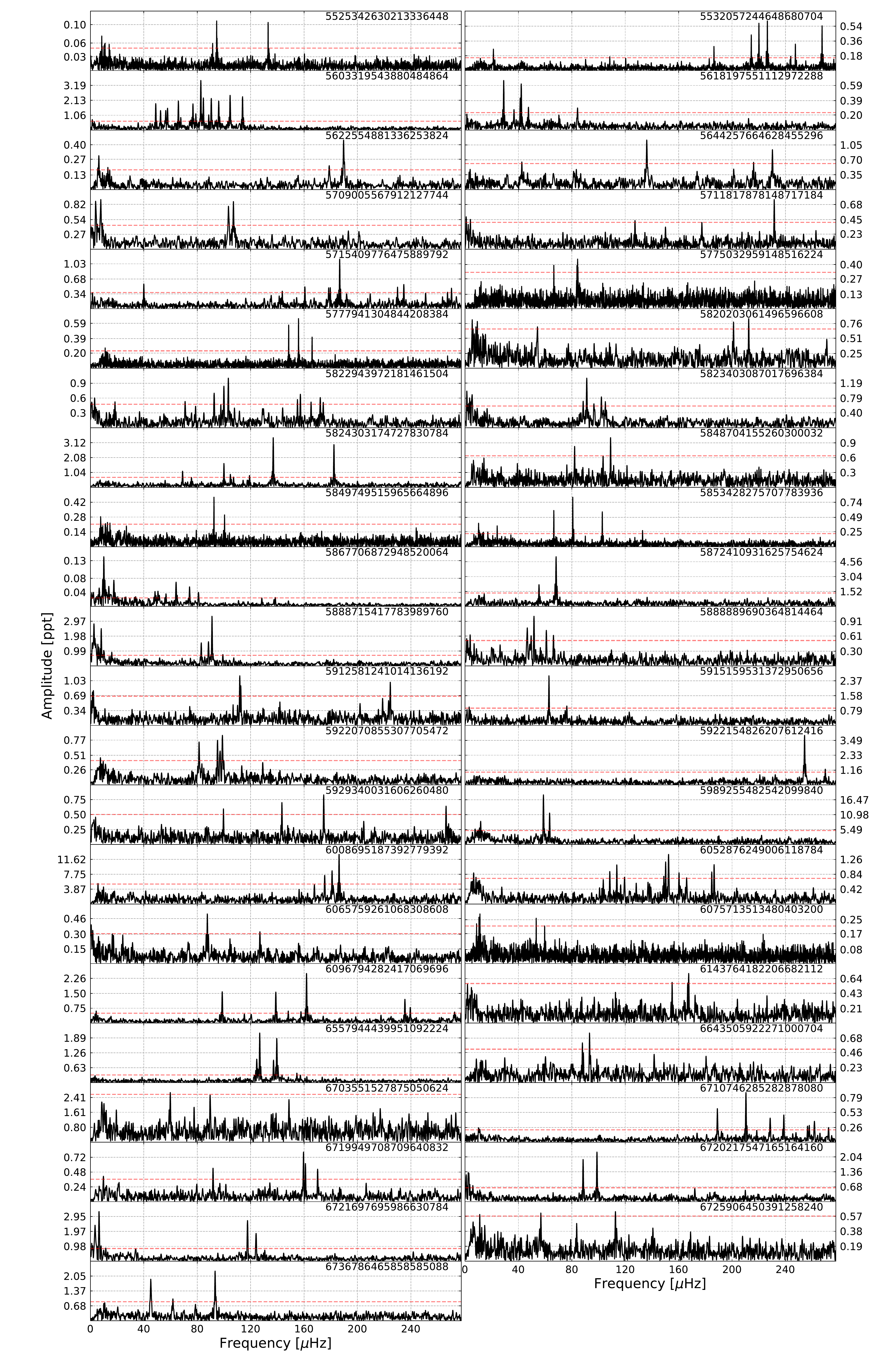}
\caption{Same as Figure\,\ref{fig:pulsators1}}
\label{fig:pulsators2}
\end{figure*}

We found 83 additional objects that show multi-peak amplitude spectra that we interpret as pulsation. These objects are not classified so we cannot make any definite conclusion on their pulsation nature. We listed these objects along with their basic information in Table\,\ref{tab:table2}. Objects in this list have at least two peaks in their amplitude spectra that are not related. Some of the objects show peaks at frequencies too low compared to known g-mode sdBVs ({\it e.g.} {\it Gaia} DR2\,5618197551112972288), unless they are very cool sdBVs, which have g-modes shifted to longer periods. We show the amplitude spectra of these pulsators in Figures\,\ref{fig:pulsators1} and \ref{fig:pulsators2}.

Even though we have no spectral classifications for these pulsator candidates, we selected five objects for preliminary mode identification. We chose objects that are rich in high amplitude peaks in a frequency range that is typical of g-modes in sdBVs. These stars may also be $\delta$\,Scuti or $\beta$\,Cep stars, however \citet{geier19} and \citet{geier20} applied a color index criterium to avoid cool stars. These papers provide detailed arguments using \textit{Gaia} color indices that these targets are hot subluminous stars and occupy the region $-0.7 < G_{BP}-G_{RP} \lesssim 0.7$ in the \textit{Gaia} colour space. Selecting only the targets which are rich in high amplitude peaks are necessary to search for multiplets and equally spaced overtones. The frequencies cannot be too close to the Nyquist frequency, since we are unable to discern between true peaks and their reflections across the Nyquist frequency. This constrains our selection to cool sdBVs, since only these objects have peaks in their amplitude spectra shifted to lower frequencies, as compared with hotter g-mode sdBVs. Even though we are not sure the selected objects are sdBVs, our suggested mode assignment may be useful if a spectral classification is confirmed by future spectroscopic analyses. We show the result of our mode identification in Section\,\ref{modeid}.

\begin{table}
	\centering
	\caption{Basic information of 83 pulsator candidates we found that are not spectroscopically classified. We show amplitude spectra of these objects in Figs.\,\ref{fig:pulsators1} and \ref{fig:pulsators2}}
	\label{tab:table2}
	\resizebox{\columnwidth}{!}{\begin{tabular}{lllll}
		\hline
		{\it Gaia} DR2 & TIC & G mag & Sector & Remarks\\
		\hline
2921500461998485248 & 744231977 & 18.3131 & 6-7 & 2 objects within 21"\\
2927637764107094272 & 744958933 & 18.6743 & 7 & crowded field\\
3062196993541803904 & 754827446 & 17.4488 & 7 & 2 objects within 21"\\
3087146252404755584 & 257068255 & 15.0790 & 7 & bright object 18" away\\
3111790534231122944 & 284329074 & 15.2648 & 7 & -\\
3135760162593729792 & 318043125 & 15.1359 & 7 & -\\
3167805630845394432 & 762068195 & 18.6149 & 7 & 3 bright objects within 31"\\
3229067399384875776 & 457225725 & 14.3108 & 5 & -\\
3343211511104992512 & 247871256 & 12.0741 & 6 & -\\
4923853724788504192 & 201251043 & 11.9467 & 1-2 & -\\
5195370596485039360 & 309437982 & 11.6366 & 11-13 & -\\
5196271513123121152 & 323174439 & 13.3202 & 11-13 & -\\
5200393440479943808 & 356730219 & 12.0231 & 12 & 2 objects within 21"\\
5224191377530306688 & 454961165 & 16.1859 & 11-12 & 2 objects within 20"\\
5233348350879239680 & 906337576 & 18.9777 & 11-12 & crowded field\\
5233397274851278720 & 906364997 & 17.9977 & 11-12 & -\\
5237830853646781440 & 910533311 & 17.9658 & 10-11 & -\\
5249563295545862400 & 846723312 & 17.9838 & 9-11 & crowded field\\
5249647030230264960 & 846766924 & 18.0805 & 9-11 & 3 objects within 21"\\
5250674622612902912 & 362098036 & 11.4953 & 9-11 & -\\
5257667310412799104 & 854453711 & 15.6269 & 9-10 & bright object 19" away\\
5257747299878049152 & 442128473 & 10.7712 & 9-10 & -\\
5266133451162548864 & 141684783 & 14.5486 & 1-13 & -\\
5291746162211858688 & 349477778 & 15.4137 & 1-13 & -\\
5298504894555743488 & 359056669 & 13.4350 & 9-11 & -\\
5311214355632943744 & 810530414 & 17.3119 & 8-10 & crowded field\\
5314453718731068288 & 811223439 & 18.1562 & 8-10 & -\\
5315487775054631040 & 811466418 & 17.6162 & 8-10 & -\\
5333948923192853376 & 281555312 & 16.1262 & 10-11 & crowded field\\
5342621080656522880 & 265631178 & 14.4456 & 10-11 & crowded field\\
5347816273079220096 & 933707513 & 16.8892 & 10 & 2 bright objects within 21"\\
5362558250096941056 & 178621334 & 13.3439 & 10 & -\\
5365175740610237568 & 864799220 & 18.6240 & 9-10 & bright object 19" away\\
5366164442080271232 & 146721954 & 16.2229 & 9-10 & CV \citep{Pretorius08}\\
5391470732981138816 & 147136095 & 15.1171 & 9-10 & -\\
5430557649797803264 & 3120302 & 16.4812 & 8-9 & bright object 11" away\\
5439887654492064256 & 25836205 & 13.1569 & 8-9 & 3 objects within 20"\\
5478585589003785088 & 737275640 & 18.4758 & 1-13 & bright object 15" away\\
5511191365810066176 & 123027362 & 15.9615 & 7-8 & 3 objects within 25"\\
5518789334757310208 & 818321152 & 18.1986 & 7-9 & very crowded field\\
5525342630213336448 & 181142865 & 11.1041 & 8-9 & -\\
5532057244648680704 & 768899830 & 18.9053 & 7-8 & very crowded field\\
5603319543880484864 & 98487756 & 14.2329 & 7 & -\\
5618197551112972288 & 100472259 & 11.5248 & 7 & TYC\,6537-2358-1 17" away\\
5622554881336253824 & 190720627 & 11.0249 & 8 & 3 objects within 21"\\
5644257664628455296 & 147520373 & 16.3055 & 8 & 2 bright objects nearby\\
5709005567912127744 & 834530823 & 16.5010 & 8 & 2 bright objects within 19"\\
5711817878148717184 & 140753138 & 12.8517 & 7 & objects within 22"\\
5715409776475889792 & 415273200 & 13.9153 & 7 & 3 objects within 21"\\
5775032959148516224 & 1205151714 & 18.5283 & 12-13 & 3 objects within 21"\\
5777941304844208384 & 310102142 & 15.9065 & 12-13 & object 24" away\\
5820203061496596608 & 1108611714 & 17.5718 & 12 & -\\
5822943972181461504 & 1109489759 & 18.2134 & 12 & crowded field\\
5823403087017696384 & 446919722 & 12.9835 & 12 & -\\
5824303174727830784 & 1110689652 & 17.1889 & 12 & bright object 16" away\\
5848704155260300032 & 1121956723 & 17.7436 & 12 & 3 objects within 21"\\
5849749515965664896 & 398597669 & 16.0006 & 11-12 & very crowded field\\
5853428275707783936 & 1019266145 & 16.3022 & 11-12 & very crowded field\\
5867706872948520064 & 1028442563 & 18.4981 & 11 & 2 objects within 21"\\
5872410931625754624 & 1035773311 & 17.5023 & 11 & 3 bright objects within 21"\\
5888715417783989760 & 461920379 & 13.9432 & 12 & object 16" away\\
5888889690364814464 & 144582845 & 15.4221 & 12 & -\\
5912581241014136192 & 360220395 & 13.8772 & 12 & GSC 08740-00359 22" away\\
5915159531372950656 & 1314011445 & 18.7489 & 12 & crowded field\\
5922070855307705472 & 388832080 & 12.5940 & 12 & -\\
5922154826207612416 & 1514267365 & 18.9560 & 12 & very crowded field\\
5929340031606260480 & 1321621063 & 18.6969 & 12 & crowded field\\
5989255482542099840 & 1163132002 & 17.9496 & 12 & crowded field\\
6008695187392779392 & 1169856895 & 18.9471 & 12 & crowded field\\
6052876249006118784 & 973849292 & 17.1040 & 11 & crowded field\\
6065759261068308608 & 1046158353 & 17.0814 & 11 & very crowded field\\
6075713513480403200 & 991374485 & 18.5056 & 10-11 & crowded field\\
6096794282417069696 & 1051339091 & 17.9689 & 11 & 3 objects within 21"\\
6143764182206682112 & 22217594 & 15.1797 & 10 & -\\
6557944439951092224 & 2027194271 & 18.6307 & 1 & overlapped with a galaxy\\
6643505922271000704 & 230975415 & 15.6128 & 13 & object 15" away\\
6703551527875050624 & 1692650109 & 17.9725 & 13 & -\\
6710746285282878080 & 1819391406 & 18.2905 & 13 & crowded field\\
6719949708709640832 & 1695487554 & 18.3141 & 13 & crowded field\\
6720217547165164160 & 1695657786 & 17.1742 & 13 & 2 bright objects within 21"\\
6721697695986630784 & 89907858 & 14.7930 & 13 & -\\
6725906450391258240 & 1701013661 & 18.3070 & 13 & very crowded field\\
6736786465858585088 & 1707873842 & 17.0868 & 13 & crowded field\\
        \hline
	\end{tabular}}
\end{table}

\subsection{Eclipsing binaries}
\begin{figure*}
\includegraphics[width=0.85\textwidth]{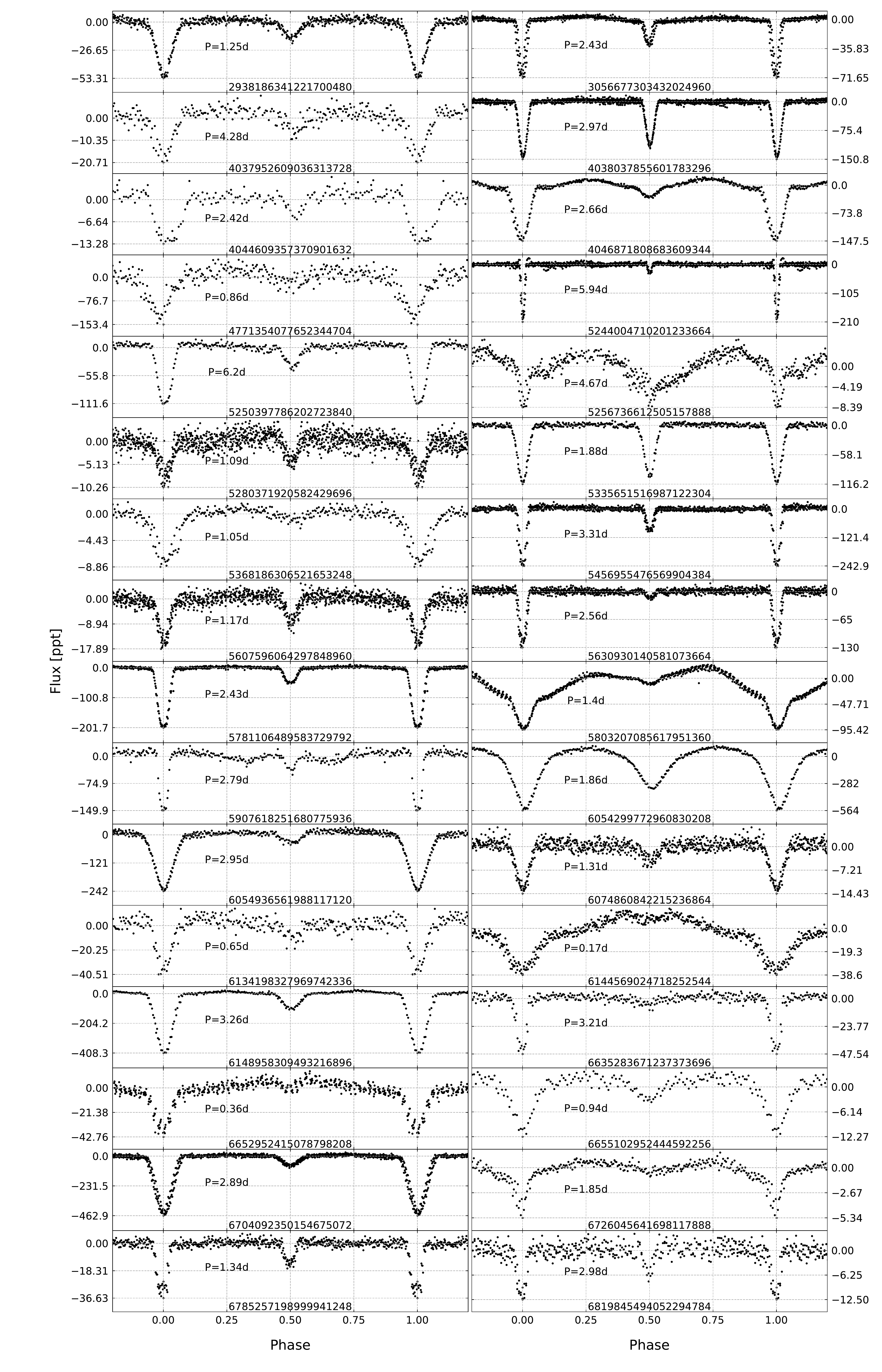}
\caption{Phased time-series data of eclipsing binaries that show both primary and secondary eclipses and are likely detached or semi-detached binaries.}
\label{fig:eb_ps}
\end{figure*}

We have found an additional 83 eclipsing binaries in our analysis that are not yet classified. They all show distinct eclipses and are likely either detached or semi-detached binaries. Possible contact binaries are not included in this group. We separated these 83 objects into three groups. 32 objects show both primary and secondary eclipses and we list them along with their basic information in Table\,\ref{tab:table3}. We show phased time-series data of these objects in Figure\,\ref{fig:eb_ps}. Two objects, {\it Gaia} DR2\,6144569024718252544 and {\it Gaia} DR2\,6652952415078798208, show additional reflection effect, which allows us to classify them as HW\,Vir systems. The latter has been already reported by \citet{drake17}.

\begin{figure*}
\includegraphics[width=0.85\textwidth]{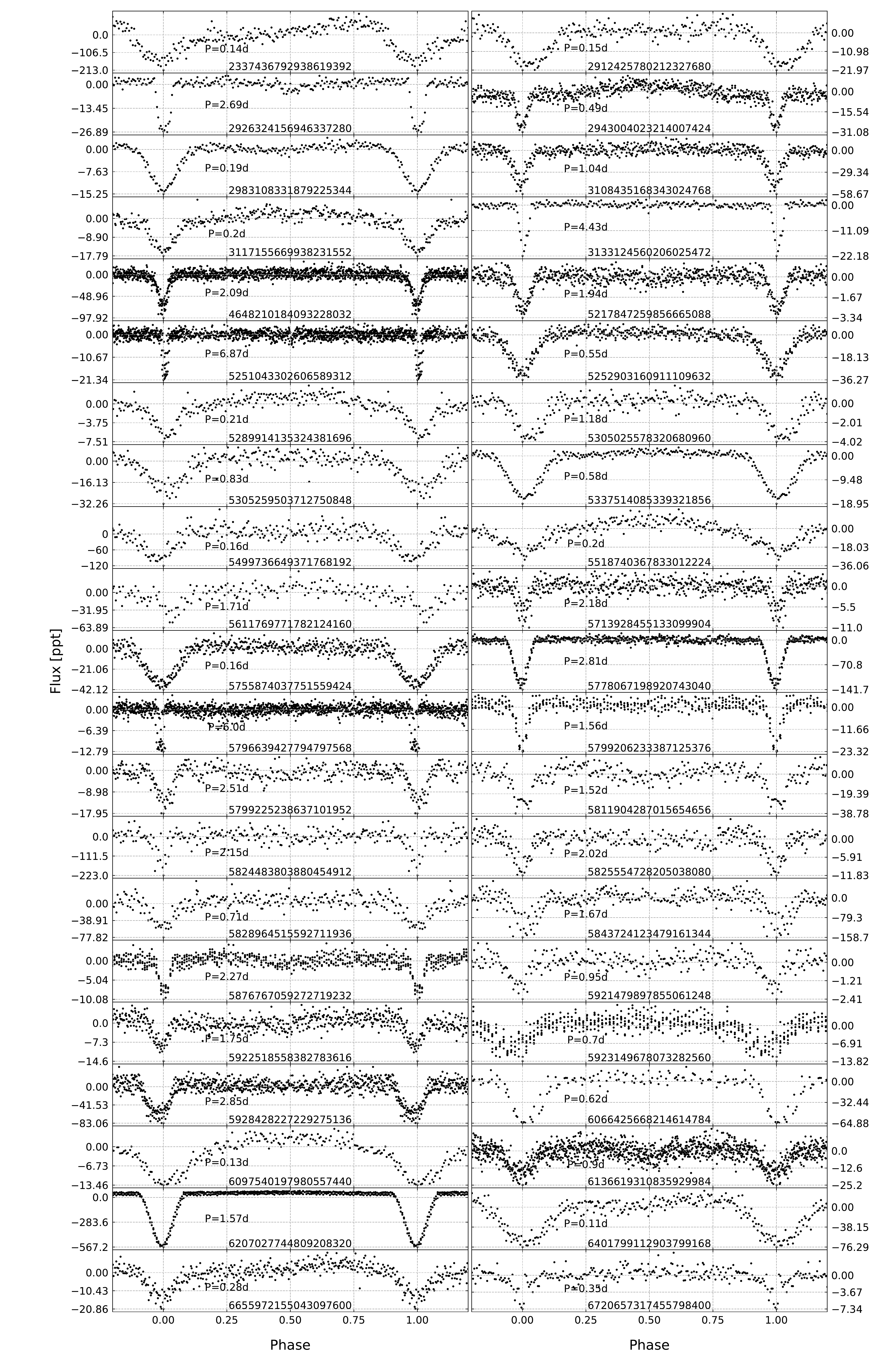}
\caption{Phased time-series data of eclipsing binaries that show only primary eclipses.}
\label{fig:eb_pr}
\end{figure*}

42 objects show only one eclipse (likely primary). The secondary eclipses are not detected in our data. This may be a consequence of a low inclination and/or small size of either companion with respect to the distance between them. These objects are listed in Table\,\ref{tab:table4} and we show phased time-series data in Figure\,\ref{fig:eb_pr}. There are three candidates for HW\,Vir systems in this sample. They show no detectable secondary eclipses but they show a flux increase between primary eclipses that is characteristic of a reflection effect. These objects are {\it Gaia} DR2\,2943004023214007424, {\it Gaia} DR2\,3117155669938231552 and {\it Gaia} DR2\,5289914135324381696. Two objects, {\it Gaia} DR2\,5518740367833012224 and {\it Gaia} DR2\,6097540197980557440, show the flux increase between primary eclipses, however the eclipses itself seem to be too wide for a compact primary component.

\begin{figure*}
\includegraphics[width=0.85\textwidth]{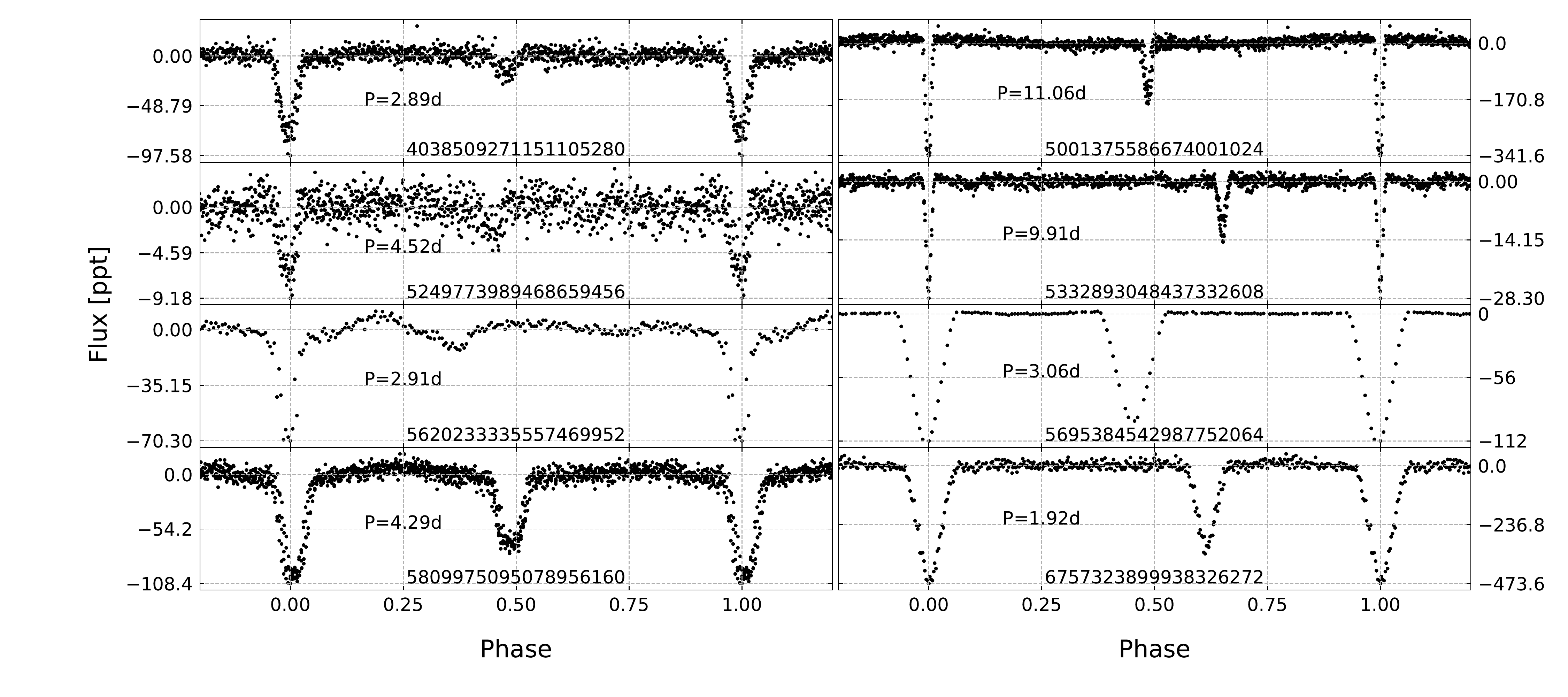}
\caption{Phased/binned time-series data of eclipsing binaries with eccentric orbits.}
\label{fig:eb_e}
\end{figure*}

\begin{figure}
\includegraphics[width=\columnwidth]{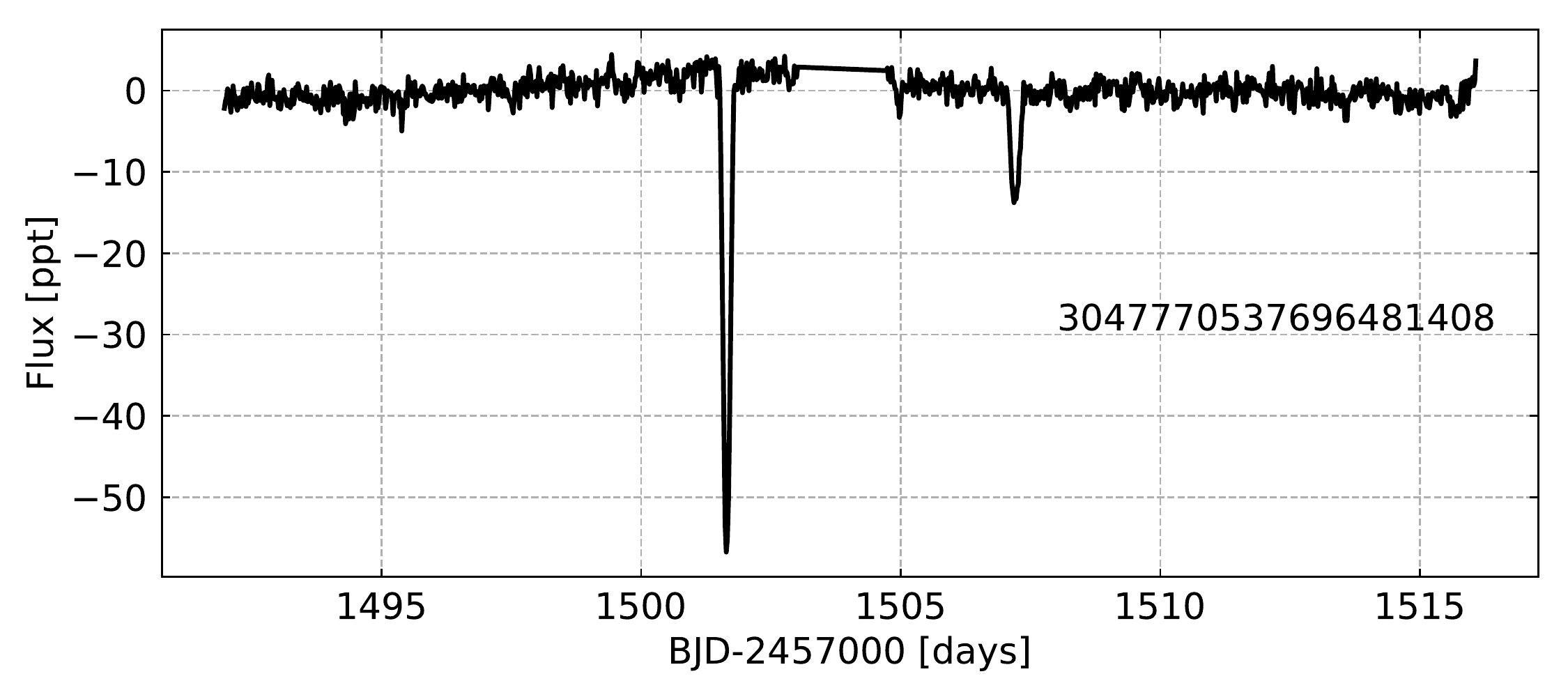}
\caption{Time-series data of eclipsing binary {\it Gaia}\,DR2\,3047770537696481408 with an eccentric orbit.}
\label{fig:eb_ec}
\end{figure}

We also found nine eclipsing binaries with secondary eclipses not centered at 0.5 phase, which we interpret as the signature of an eccentric orbit. We list these objects in Table\,\ref{tab:table5} and we show phased time-series data in Figure\,\ref{fig:eb_e}. In Figure\,\ref{fig:eb_ec} we present one object which is likely a binary with eccentric orbit but we have not detected either two primary or secondary eclipses and consequently we are unable to determine an orbital period. If the system were not eccentric we would see the consecutive primary eclipse at around BJD-2457000\,=\,1512.5.

\begin{table*}
	\centering
	\caption{Basic information of 32 eclipsing binaries that show both primary and secondary eclipses. We show the time-series data of these objects in Figure\,\ref{fig:eb_ps}}
	\label{tab:table3}
	\resizebox{0.85\textwidth}{!}{\begin{tabular}{llllll}
		\hline
		{\it Gaia} DR2 & TIC & G mag & Sector & Period[d] & Remarks\\
		\hline
2938186341221700480 & 60523137 & 16.2325 & 6 & 1.2532 & 3 bright objects within 21"\\
3056677303432024960 & 753916356 & 17.9663 & 7 & 2.4282 & 3 objects within 21"\\
4037952609036313728 & 1556986400 & 18.8581 & 13 & 4.2844 & very crowded field\\
4038037855601783296 & 1557298522 & 17.1815 & 13 & 2.9735 & 2 objects within 17"\\
4044609357370901632 & 1569961982 & 16.4584 & 13 & 2.415 & very crowded field\\
4046871808683609344 & 1577179385 & 17.9382 & 13 & 2.6564 & crowded field\\
4771354077652344704 & 685209944 & 17.0220 & 3-6,10,13 & 0.8619 & -\\
5244004710201233664 & 371323868 & 16.7718 & 10,11 & 5.9367 & 2 bright objects within 30"\\
5250397786202723840 & 847157315 & 17.3345 & 9-11 & 6.1989 & crowded field\\
5256736612505157888 & 462454783 & 13.0236 & 9-10 & 4.6739 & -\\
5280371920582429696 & 176935277 & 15.8967 & 1-13 & 1.0922 & -\\
5335651516987122304 & 321175529 & 12.1788 & 10-11 & 1.8844 & crowded field\\
5368186306521653248 & 146317976 & 14.3942 & 9-10 & 1.0496 & crowded field\\
5456955476569904384 & 942871718 & 18.0126 & 9 & 3.3063 & very bright star 9" away\\
5607596064297848960 & 778203221 & 17.2286 & 7 & 1.1652 & -\\
5630930140581073664 & 827970537 & 17.0919 & 8,9 & 2.5561 & 3 bright objects within 20"\\
5781106489583729792 & 309959661 & 16.1617 & 12-13 & 2.4251 & 2 bright objects within 20"\\
5803207085617951360 & 1509000006 & 16.6916 & 12-13 & 1.3974 & object 14" away\\
5907618251680775936 & 1155805356 & 18.0498 & 11 & 2.7874 & a bright star is at 12"\\
6054299772960830208 & 976146222 & 16.7204 & 10-11 & 1.8626 & very crowded field\\
6054936561988117120 & 450420664 & 10.6360 & 11 & 2.9541 & \citet{Avvakumova13}\\
6074860842215236864 & 990883707 & 18.0659 & 11 & 1.3087 & 2 objects within 21"\\
6134198327969742336 & 996495522 & 17.9314 & 10-11 & 0.6491 & -\\
6144569024718252544 & 258379678 & 15.3605 & 10 & 0.1739 & \citet{drake17}; HW\,Vir\\
6148958309493216896 & 998016645 & 18.0266 & 10 & 3.2584 & 3 bright objects within 21"\\
6635283671237373696 & 1689813971 & 18.9809 & 13 & 3.2146 & 2 bright objects within 21"\\
6652952415078798208 & 76760933 & 13.8504 & 13 & 0.3613 & HW\,Vir candidate\\
6655102952444592256 & 1817787859 & 16.5162 & 13 & 0.9444 & -\\
6704092350154675072 & 1692824895 & 16.7411 & 13 & 2.8881 & CRTS\,J183755.6-484955 \citep{drake17}\\
6726045641698117888 & 1701292115 & 9.1796 & 13 & 1.8459 & 3 objects nearby\\
6785257198999941248 & 270535918 & 13.4223 & 1 & 1.3423 & -\\
6819845494052294784 & 2028782100 & 18.2564 & 1 & 2.9797 & -\\
        \hline
	\end{tabular}}
\end{table*}

\begin{table*}
	\centering
	\caption{Basic information of 42 eclipsing binaries that show primary eclipses only. We show the time-series data of these objects in Figure\,\ref{fig:eb_pr}.}
	\label{tab:table4}
	\resizebox{0.85\textwidth}{!}{\begin{tabular}{llllll}
		\hline
		{\it Gaia} DR2 & TIC & G mag & Sector & Period[d] & Remarks\\
		\hline
2337436792938619392 & 33984762 & 15.4745 & 2 & 0.1445 & Nova \citep{samus17}\\
2912425780212327680 & 37737816 & 16.9636 & 6 & 0.1488 & -\\
2926324156946337280 & 707111651 & 17.8191 & 6-7 & 2.6908 & object 10" away\\
2943004023214007424 & 33743252 & 14.0155 & 6 & 0.4886 & 2 objects within 25"; HW\,Vir candidate\\
2983108331879225344 & 189012795 & 15.3166 & 5 & 0.1889 & object 13" away\\
3108435168343024768 & 756875938 & 17.2058 & 7 & 1.0354 & 2 bright objects within 21"\\
3117155669938231552 & 42566802 & 16.0394 & 6 & 0.1986 & HW\,Vir candidate\\
3133124560206025472 & 202273662 & 12.3871 & 6 & 4.4319 & 2 objects within 21"\\
4648210184093228032 & 141280240 & 15.7999 & 1-2,4-12 & 2.0908 & -\\
5217847259856665088 & 843283217 & 17.9546 & 10-12 & 1.9444 & object 16" away\\
5251043302606589312 & 847473488 & 18.2467 & 9-10 & 6.8704 & 3 objects within 21"\\
5252903160911109632 & 849266771 & 18.8591 & 10-11 & 0.5458 & crowded field\\
5289914135324381696 & 308541002 & 16.5557 & 1,4,7-11 & 0.2105 & HW\,Vir candidate\\
5305025578320680960 & 383375636 & 15.5078 & 8-10 & 1.1821 & crowded field\\
5305259503712750848 & 856066667 & 18.4776 & 9 & 0.8298 & 4 bright objects within 21"\\
5337514085339321856 & 280246753 & 12.2093 & 10-11 & 0.5821 & crowded field\\
5499736649371768192 & 255594396 & 17.4141 & 1-2,5-9,11-12 & 0.1585 & -\\
5518740367833012224 & 818308005 & 17.4758 & 7-8,9 & 0.2049 & 2 bright objects within 21"\\
5611769771782124160 & 779128665 & 18.9927 & 7 & 1.7065 & 4 brighter objects within 21"\\
5713928455133099904 & 834924649 & 18.3591 & 7 & 2.1812 & 2 objects within 21"\\
5755874037751559424 & 95785714 & 15.8131 & 8 & 0.1559 & -\\
5778067198920743040 & 1309769412 & 17.8042 & 12-13 & 2.8099 & object 5" away\\
5796639427794797568 & 401617089 & 13.1231 & 12 & 5.9992 & 3 objects within 26"\\
5799206233387125376 & 1106932458 & 18.1909 & 12 & 1.5615 & crowded field\\
5799225238637101952 & 1106947173 & 18.2872 & 12 & 2.5142 & -\\
5811904287015654656 & 1509561926 & 17.7235 & 12-13 & 1.5239 & 2 objects within 21"\\
5824483803880454912 & 1110889969 & 17.7445 & 12 & 2.1506 & crowded field\\
5825554728205038080 & 424717152 & 15.0860 & 12 & 2.0156 & bright object 9" away\\
5828964515592711936 & 1209068896 & 17.3298 & 12 & 0.7066 & 5 objects within 21"\\
5843724123479161344 & 957158511 & 17.2414 & 11-12 & 1.673 & crowded field\\
5876767059272719232 & 1130777039 & 15.8884 & 12 & 2.2711 & crowded field\\
5921479897855061248 & 1513749999 & 18.3774 & 12 & 0.9524 & crowded field\\
5922518558382783616 & 1315988153 & 18.8932 & 12 & 1.7541 & 3 bright objects within 21"\\
5923149678073282560 & 1316233834 & 18.4008 & 12 & 0.698 & 3 bright objects within 21"\\
5928428227229275136 & 1219774587 & 18.9003 & 12 & 2.8496 & crowded field\\
6066425668214614784 & 986927321 & 17.3006 & 11 & 0.6241 & very crowded field\\
6097540197980557440 & 242402846 & 16.3590 & 11 & 0.1274 & -\\
6136619310835929984 & 996840740 & 18.3096 & 11 & 0.9006 & object 14" away\\
6207027744809208320 & 1174994870 & 17.7766 & 11 & 1.5731 & -\\
6401799112903799168 & 410442849 & 14.6995 & 13 & 0.1117 & CV \citep{ODonoghue13}\\
6655972155043097600 & 456516418 & 13.5608 & 13 & 0.2786 & -\\
6720657317455798400 & 1695999069 & 16.2634 & 13 & 0.3536 & crowded field\\
        \hline
	\end{tabular}}
\end{table*}

\begin{table*}
	\centering
	\caption{Basic information of nine eclipsing binaries with eccentric orbits. We show the time-series data of these objects in Figures\,\ref{fig:eb_e} and \ref{fig:eb_ec}}
	\label{tab:table5}
	\resizebox{0.85\textwidth}{!}{\begin{tabular}{llllll}
		\hline
		{\it Gaia} DR2 & TIC & G mag & Sector & Period[d] & Remarks\\
		\hline
3047770537696481408 & 187657618 & 9.8610 & 7 & - & TYC\,5395-2586-2 nearby\\
4038509271151105280 & 1558593400 & 18.7875 & 13 & 2.8936 & very crowded field\\
5001375586674001024 & 616535392 & 18.7768 & 2 & 11.0648 & -\\
5249773989468659456 & 846855557 & 18.4170 & 9-11 & 4.5177 & 2 objects within 21"\\
5332893048437332608 & 305898260 & 11.8255 & 10-11 & 9.9141 & crowded field\\
5620233335557469952 & 139334900 & 13.0033 & 7 & 2.9054 & -\\
5695384542987752064 & 832426099 &  9.8621 & 8 & 3.0607 & AC\,Pyx \citep{samus03}\\
5809975095078956160 & 1509255862 & 18.9775 & 12-13 & 4.2926 & 2 bright objects within 21"\\
6757323899938326272 & 1826847776 & 18.8560 & 13 & 1.9224 & 3 bright objects within 21"\\
        \hline
	\end{tabular}}
\end{table*}

\subsection{Spectroscopically unclassified variables}
We found a spectroscopically unidentified sample of 1536 variable stars that we identified neither as pulsators nor as eclipsing binary systems, presented in the previous subsections. The proper variability type can only be done with a spectroscopic classification and radial velocity curves. We split this sample into two groups.

\begin{figure*}
\includegraphics[width=0.85\textwidth]{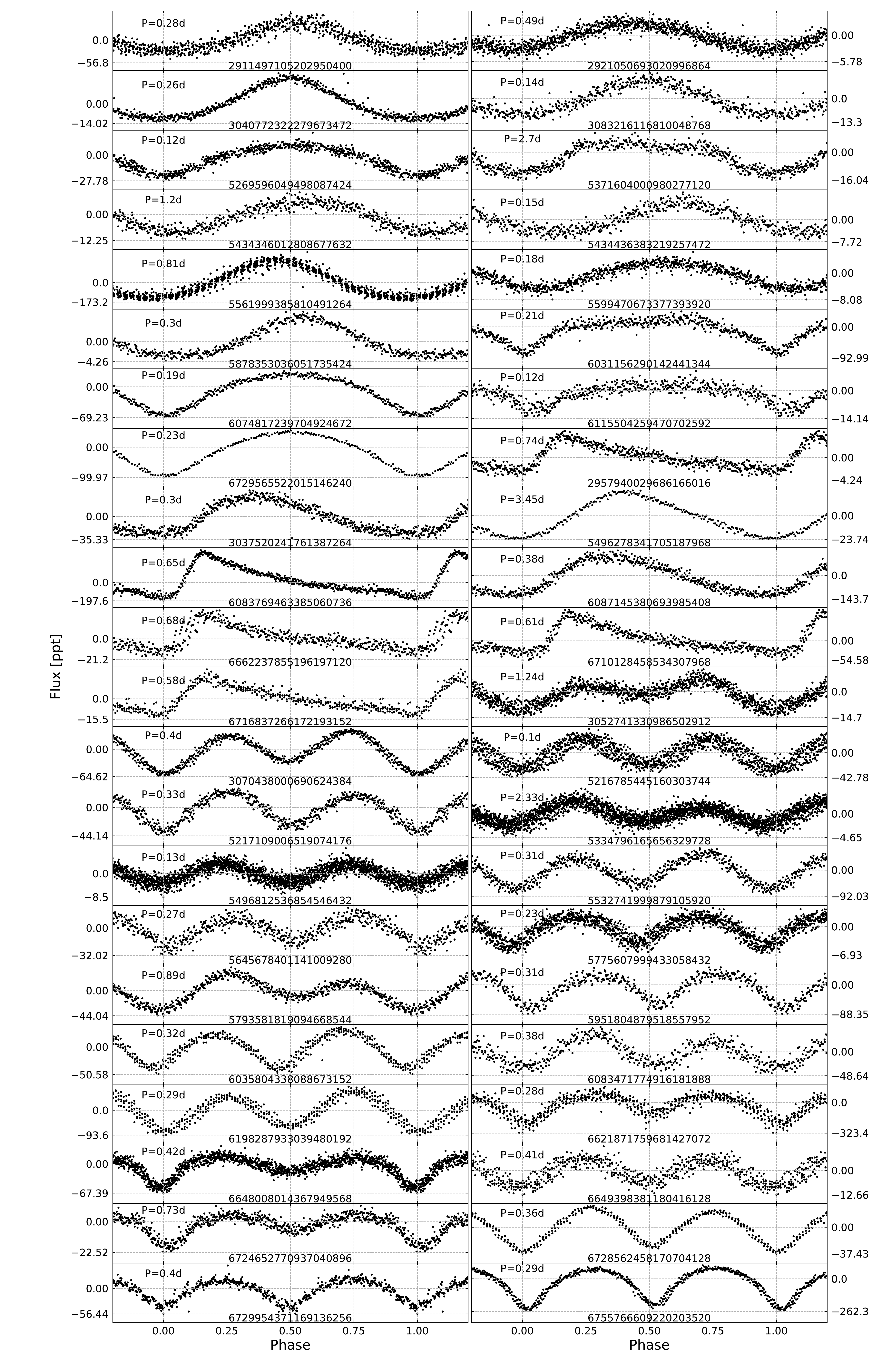}
\caption{Phased time-series data for variable objects that are spectroscopically unclassified. The first fifteen objects have one symmetric maximum, the next eight objects have one asymmetric maximum, and the final twenty-one objects have two maxima.}
\label{fig:nospec-ph}
\end{figure*}

The first group contains targets that show peaks consistent with binarity, and amplitudes of a flux variation that is large enough to be clearly seen in phased data. The typical S/N is 45. The phased time-series data of these objects show either one maximum or two maxima of a flux variation. Those with one maximum can be symmetric or asymmetric. The symmetric case can be interpreted as a reflection binary \citep{baran19}. The asymmetric case is characteristic of some of the classical pulsators, {\it e.g.} RR\,Lyrae stars, anomalous Cepheids, classical Cepheids. Two maxima cases can be explained by ellipsoidal variables or contact eclipsing binaries {\it e.g.} W\,UMa stars. If the maxima are not equal, it may indicate a Doppler boosting contribution (former case), which can even end up with just one maximum mimicking a reflection binary \citep{reed16}, or the O'Connell effect (latter case). Examples of all these three cases selected in our sample are plotted in Figure\,\ref{fig:nospec-ph}. We provided basic information about these targets in Table\,\ref{tab:table6}, while the data are plotted in Figure\,\ref{fig:nospec-ph}.

\begin{figure*}
\includegraphics[width=0.85\textwidth]{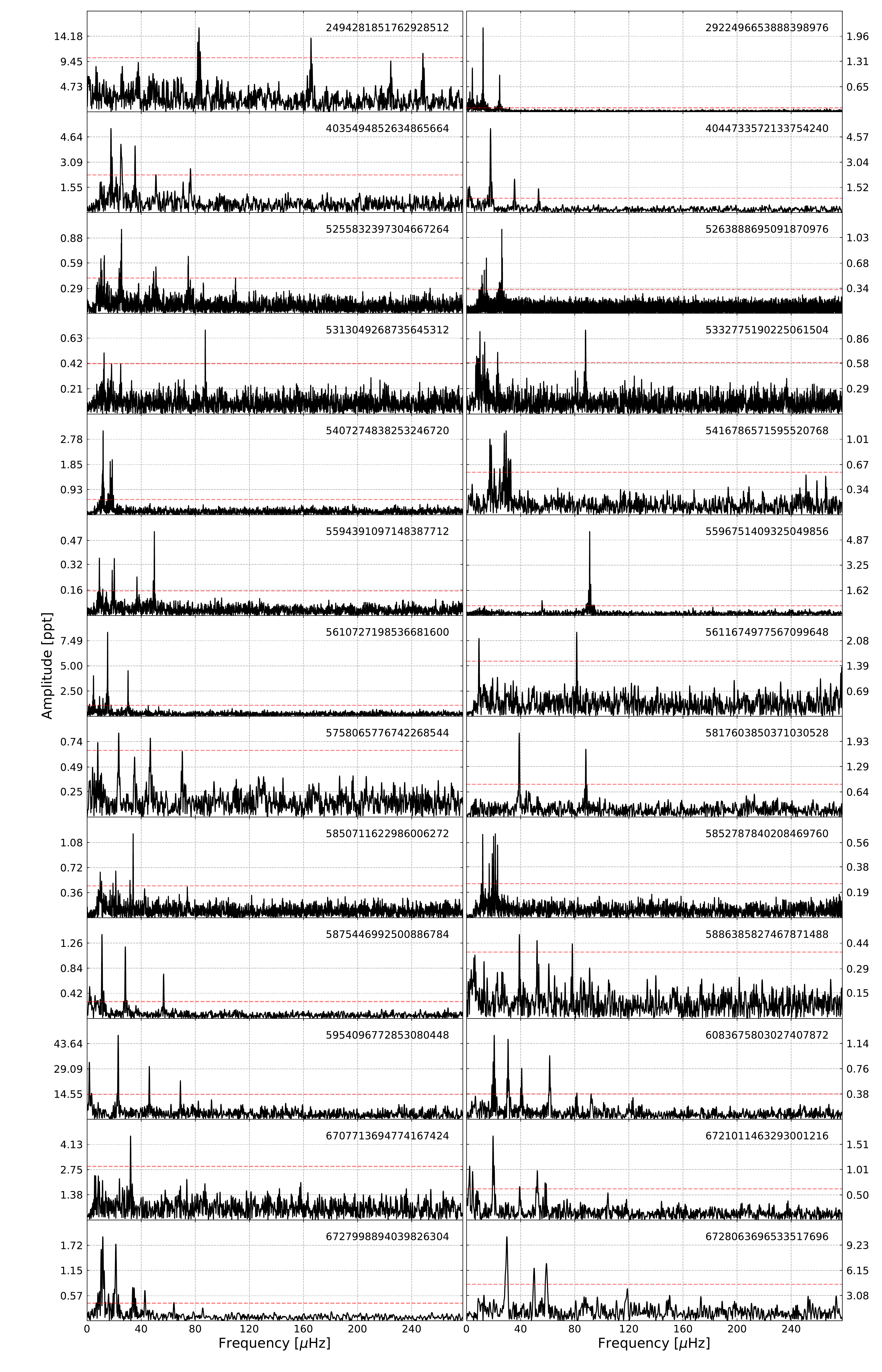}
\caption{Amplitude spectra of variable objects that are spectroscopically unclassified.}
\label{fig:nospec-fr}
\end{figure*}

The second group contains targets that show peaks consistent with binarity, and flux variation amplitudes that are too small to be clearly seen in phased data. The typical S/N is 8. We find these variations in amplitude spectra. This group contains also targets with multipeak amplitude spectra, regardless of the S/N, which makes data phasing unnecessary. This multiplicity of peaks is characteristic of pulsators, however the amplitude spectra do not resemble the ones of sdBVs, since the unrelated peaks are below 60\,$\upmu$Hz, and that is why we decided not to include them in Section\,\ref{pulsators}. We found flux variations in targets of this group in their amplitude spectra. We provided basic information about these targets in Table\,\ref{tab:table7}, while the data are plotted in Figure\,\ref{fig:nospec-fr}.

\begin{figure}
\includegraphics[width=\columnwidth]{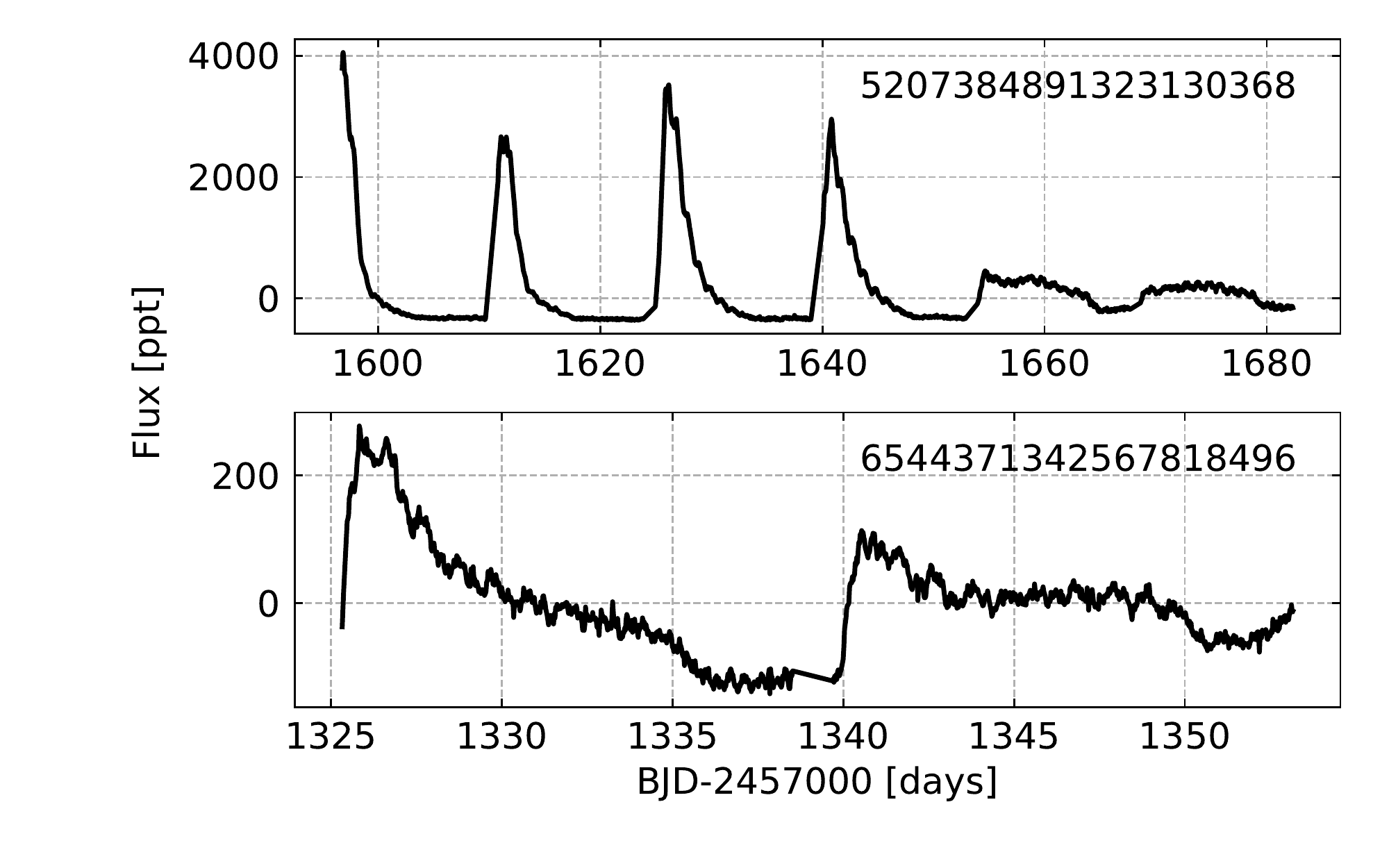}
\caption{Time-series data during outbursting of two Nova stars.}
\label{fig:nospec-lc}
\end{figure}

We have also detected two Nova stars, {\it Gaia} DR2\,5207384891323130368 and {\it Gaia} DR2\,6544371342567818496, during outbursts. The latter star is spectroscopically classified and formally included in the subsequent subsection. We show the time-series data in Figure\,\ref{fig:nospec-lc}. Both stars have been known before but {\it TESS} data recorded outbursts and that is why we included these objects in our list. More information on these two objects can be find in Table\,\ref{tab:table8}. Other known Nova detected in our sample are not specifically mentioned, since time-series data of these objects do not show significant outbursts.

\begin{table}
	\centering
	\caption{Basic information of 44 (out of 273) spectroscopically unclassified variables. The top section lists 15 objects that show one symmetric maximum, the middle section lists eight objects that show one asymmetric maximum, and the bottom section lists 21 objects that show two maxima. We show phased time-series data of these objects in Figure\,\ref{fig:nospec-ph}. The list of remaining objects (229) are included in on-line material only.}
	\label{tab:table6}
	\resizebox{\columnwidth}{!}{\begin{tabular}{llllll}
		\hline
		{\it Gaia} DR2 & TIC & G mag & Sector & Period[d] & Remarks\\
		\hline
2911497105202950400 & 37004041 & 15.1572 & 6 & 0.2833 & \citet{drake17}\\
2921050693020996864 & 63113578 & 11.4451 & 6-7 & 0.4854 & object 11" away\\
3040772322279673472 & 32302937 & 14.2637 & 7 & 0.2618 & -\\
3083216116810048768 & 73238638 & 15.3182 & 7 & 0.1376 & -\\
5269596049498087424 & 765017261 & 17.1323 & 1-13 & 0.1247 & -\\
5371604000980277120 & 61762775 & 12.6910 & 10 & 2.7027 & -\\
5434346012808677632 & 70717873 & 13.6554 & 9 & 1.2048 & -\\
5434436383219257472 & 29723252 & 14.0488 & 9 & 0.1502 & -\\
5561999385810491264 & 170310610 & 15.3198 & 6-7 & 0.813 & \citet{drake17}\\
5599470673377393920 & 775878600 & 16.1312 & 7 & 0.1767 & -\\
5878353036051735424 & 1036707862 & 15.0629 & 11-12 & 0.3026 & crowded field\\
6031156290142441344 & 191221080 & 12.8298 & 12 & 0.2053 & crowded field\\
6074817239704924672 & 990861367 & 17.7796 & 11 & 0.1887 & -\\
6115504259470702592 & 166894438 & 15.8446 & 11 & 0.1238 & -\\
6729565522015146240 & 1821904001 & 18.1963 & 13 & 0.2268 & TYC\,7919-615-1 20" away\\
\hline
2957940029686166016 & 708350292 & 17.5007 & 5-6 & 0.7353 & very crowded field\\
3037520241761387264 & 750862698 & 18.9978 & 7 & 0.3003 & 5 objects within 21"\\
5496278341705187968 & 737473774 & 16.7776 & 1-13 & 3.4483 & bright object 13" away\\
6083769463385060736 & 1048223010 & 16.7149 & 11 & 0.6536 & -\\
6087145380693985408 & 1048399825 & 18.8821 & 11 & 0.3831 & -\\
6662237855196197120 & 1818520053 & 18.8787 & 13 & 0.6849 & 2 objects within 21"\\
6710128458534307968 & 1819242966 & 18.3171 & 13 & 0.6098 & 4 objects within 21"\\
6716837266172193152 & 1820684543 & 18.5151 & 13 & 0.578 & 3 objects within 21"\\
\hline
3052741330986502912 & 125197892 & 12.7196 & 7 & 1.2426 & 3 objects within 25"\\
3070438000690624384 & 803473779 & 17.6109 & 7 & 0.4016 & 3 objects within 21"\\
5216785445160303744 & 287977499 & 12.5806 & 6,10-12 & 0.0976 & \citet{Ratzloff19}\\
5217109006519074176 & 804970406 & 18.7363 & 10-12 & 0.3344 & 3 objects within 21"\\
5334796165656329728 & 325158549 & 11.0515 & 10-11 & 2.3256 & crowded field\\
5496812536854546432 & 278861557 & 15.2643 & 1-13 & 0.1318 & \citet{Kosakowski_2020}\\
5532741999879105920 & 821330826 & 17.5457 & 7-8 & 0.3077 & 3 objects within 21"\\
5645678401141009280 & 830270910 & 18.8839 & 8 & 0.2713 & 2 bright objects within 21"\\
5775607999433058432 & 1205184125 & 16.7156 & 12-13 & 0.2342 & -\\
5793581819094668544 & 1105179202 & 18.9163 & 12 & 0.8929 & 2 bright objects within 21"\\
5951804879518557952 & 1523243111 & 17.5232 & 12 & 0.3058 & very crowded field\\
6035804338088673152 & 1255310693 & 18.8790 & 12 & 0.3185 & 4 bright objects within 21"\\
6083471774916181888 & 1048022421 & 17.0149 & 11 & 0.3774 & 2 bright objects within 21"\\
6198287933039480192 & 1173764699 & 17.0215 & 11 & 0.2857 & -\\
6621871759681427072 & 2028297898 & 18.8353 & 1 & 0.2833 & -\\
6648008014367949568 & 1690227091 & 17.0039 & 13 & 0.4157 & -\\
6649398381180416128 & 119153557 & 16.3566 & 13 & 0.4132 & object 7" away\\
6724652770937040896 & 1699111897 & 17.1528 & 13 & 0.7317 & crowded field\\
6728562458170704128 & 1821458807 & 17.1410 & 13 & 0.3584 & bright object 8" away\\
6729954371169136256 & 1704724045 & 18.6622 & 13 & 0.3953 & crowded field\\
6755766609220203520 & 1826089601 & 17.7117 & 13 & 0.2933 & 2 objects within 21"\\
        \hline
	\end{tabular}}
\end{table}

\begin{table}
	\centering
	\caption{Basic information of 26 (out of 1262) spectroscopically unclassified variables. We show amplitude spectra of these objects in Figure\,\ref{fig:nospec-fr}. The figures of remaining objects (1236) are included in on-line material only.}
	\label{tab:table7}
	\resizebox{\columnwidth}{!}{\begin{tabular}{lllll}
		\hline
		{\it Gaia} DR2 & TIC & G mag & Sector & Remarks\\
		\hline
2494281851762928512 & 250416977 & 14.3723 & 4 & Nova \citep{Downes01}\\
2922496653888398976 & 744429291 & 10.7118 & 6-7 & TYC\,6522-1975-1 2" away\\
4035494852634865664 & 1552304995 & 16.7775 & 13 & crowded field\\
4044733572133754240 & 1570330815 & 18.6726 & 13 & very crowded field\\
5255832397304667264 & 852628533 & 17.7834 & 9-11 & -\\
5263888695091870976 & 271554913 & 15.9580 & 1-7,9-13 & -\\
5313049268735645312 & 299705972 & 16.2419 & 9-10 & -\\
5332775190225061504 & 321662350 & 15.4428 & 10-11 & 2 bright objects within 15"\\
5407274838253246720 & 867297951 & 17.8164 & 9-10 & crowded field\\
5416786571595520768 & 870414976 & 17.5691 & 9 & 3 objects within 21"\\
5594391097148387712 & 151005205 & 16.8940 & 7-8 & TYC\,7123-1718-1 19" away\\
5596751409325049856 & 154909544 & 15.0666 & 7-8 & 2 objects within 21"\\
5610727198536681600 & 778822592 & 16.0649 & 6-7 & object 17" away\\
5611674977567099648 & 109931573 & 15.9019 & 7 & crowded field\\
5758065776742268544 & 60659496 & 15.3353 & 8 & -\\
5817603850371030528 & 447448883 & 15.6791 & 12 & 5 bright objects within 21"\\
5850711622986006272 & 1012682181 & 16.2865 & 11-12 & crowded field\\
5852787840208469760 & 1017631564 & 16.3352 & 11-12 & crowded field\\
5875446992500886784 & 455460222 & 10.7497 & 12 & -\\
5886385827467871488 & 46197886 & 16.0325 & 12 & 4 objects within 21"\\
5954096772853080448 & 1526877224 & 17.4808 & 13 & 2 objects within 21"\\
6083675803027407872 & 1048099460 & 16.8865 & 11 & very crowded field\\
6707713694774167424 & 1694125610 & 16.6361 & 13 & object 13" away\\
6721011463293001216 & 1696341981 & 18.5560 & 13 & crowded field\\
6727998894039826304 & 1703871394 & 18.4060 & 13 & very crowded field\\
6728063696533517696 & 1703965520 & 18.9681 & 13 & crowded field\\
        \hline
	\end{tabular}}
\end{table}

\subsection{Non-sdB classified variables}
We have also pulled out fluxes of additional 77 objects that are spectroscopically classified. These were first considered candidates for hot stars, but spectral analyses confirmed them as non-sdB objects, mostly O, B or A main sequence stars. We found the same collection of a flux variation as in case of the sample of spectroscopically unclassified objects and we present it grouped the same way, {\it i.e.} pulsator candidates, eclipsing, reflection, ellipsoidal binaries, and classical pulsators. A table and figures showing the list of objects with their data are included in the on-line material only. The table includes basic information on each object along with additional references or contaminating objects, if any. The spectroscopically classified nova showing outbursts is included in Table\,\ref{tab:table8}.

\begin{table}
	\centering
	\caption{Basic information of two Nova stars. We show the time-series data in Figure\,\ref{fig:nospec-lc}. Both stars are reported by \citet{samus03}.}
	\label{tab:table8}
	\resizebox{\columnwidth}{!}{\begin{tabular}{llllll}
		\hline
		{\it Gaia} DR2 & TIC & Name & Spectral type & G mag & Sector\\
		\hline
5207384891323130368 & 735128403 & AH\,Men & - & 13.5101 & 1,4,11-13\\
6544371342567818496 & 121422158 & RZ\,Gru & B6 & 12.6299 & 1\\
        \hline
	\end{tabular}}
\end{table}

\section{Mode identification}
\label{modeid}
We selected five objects out of 83 non-classified pulsators to identify their pulsation geometry by assuming them as g-mode rich sdBVs. {\it Gaia} DR2\,IDs of these objects are 5233348350879239680, 5250674622612902912, 5365175740610237568, 5391470732981138816, 6096794282417069696. We provided arguments for our selection in Section\,\ref{pulsators}.

\begin{figure*}
\includegraphics[width=0.85\textwidth]{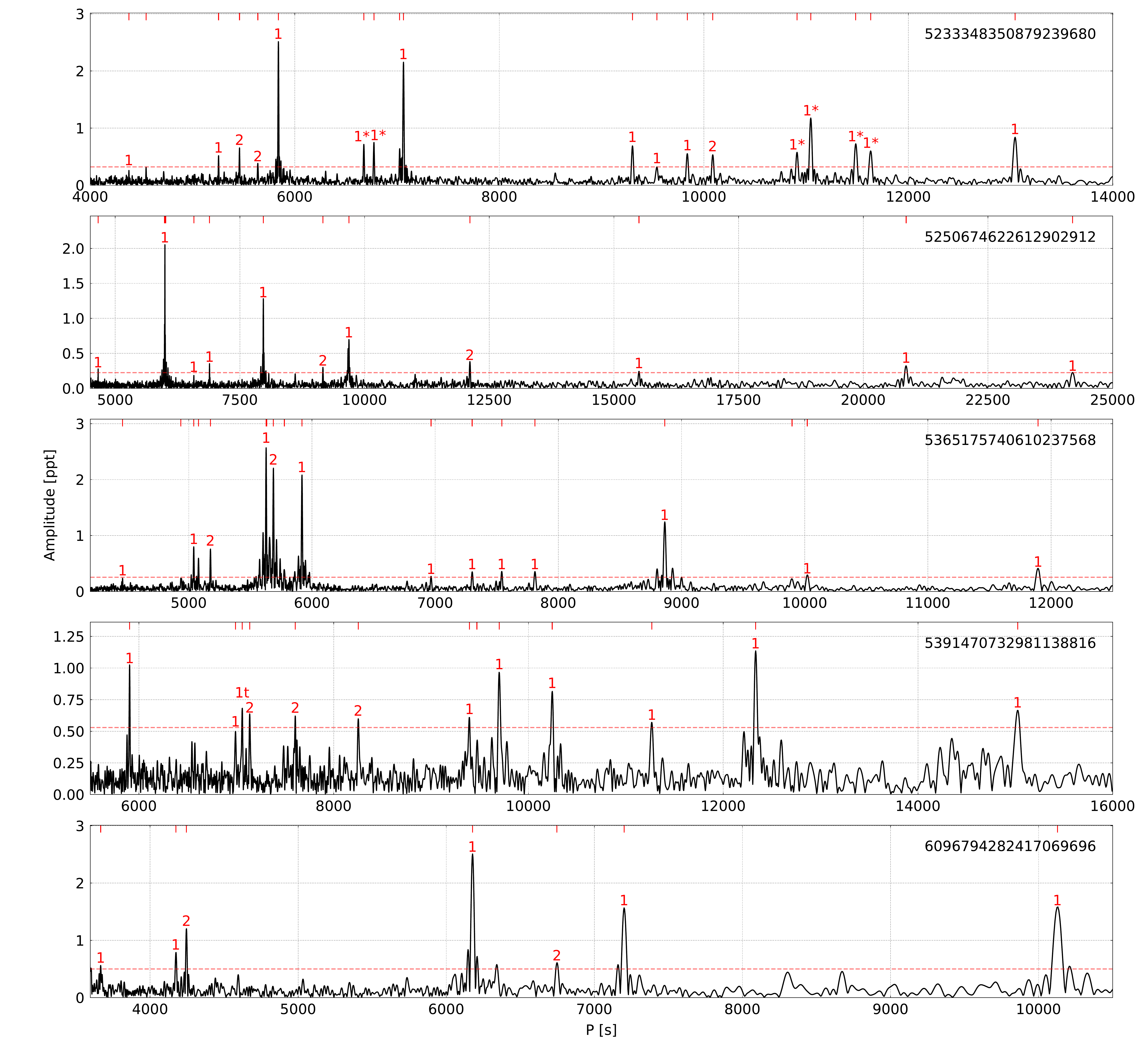}
\caption{Amplitude spectra of five pulsators plotted versus period (instead of frequency). The red horizontal dashed lines denote $4.5\,\sigma$ detection threshold. The values of modal degrees are shown on top of each detected peak.}
\label{fig:modeids}
\end{figure*}

We followed a standard prewhitening procedure by calculating an amplitude spectrum and removing consecutive peaks by fitting $A_i\sin(2{\pi}f_it+{\phi_i})$ using a non-linear least-square method, where A$_i$ is an amplitude, f$_i$ is a frequency and $\phi_i$ is a phase of an {\it i-th} peak. We used our custom scripts for prewhitening. We removed all peaks down to a detection threshold of 4.5 times the mean noise level. We updated this level after each peak removal, hence the final level was calculated from the residual amplitude spectra, {\it i.e.} with all significant peaks removed.
The lists of frequencies detected in each star are listed in Tables\,\ref{tab:table9}, \ref{tab:table10}, \ref{tab:table11}, \ref{tab:table12} and \ref{tab:table13}, while we show the amplitude spectra in Figure\,\ref{fig:modeids}.

In the case of no rotation, one cannot tell the non-radial modes of degree\,({\it l}) value of a peak, as only one peak is present. Rotation splits a mode into 2{\it l}+1 components of different {\it m} values. The frequency shift is given by the following equation $\Delta{\nu}_{n,l,m}=1-C_{n,l}/P_{\rm rot}$. According to \citet{charpinet00}, for gravity modes, the Ledoux constant C$_{n,l}$, can be calculated from the following expression $C_{n,l}=(l^2+1)^{-1}$. When the frequency shift is measured in an amplitude spectrum, a rotation period P$_{\rm rot}$ can be derived. The formulas for the frequency shift and for $C_{n,l}$ assume first order corrections and are in the asymptotic limits since the frequency range corresponds to an overtone number of 20 or greater in a typical sdBV star.

In sdBVs, in the asymptotic regime {\it i.e.} $n\gg l$, consecutive overtones of gravity modes are equally spaced in period \citep[e.g.][]{charpinet00,reed11}. Previous analyses of photometric {\it Kepler} space data of sdBVs showed that the average period spacing of dipole modes is nearly 250\,s \citep{reed18b}. The period spacings for higher degree modes can be calculated using the following relation $\Delta P_l={P_0}/{\sqrt{l(l+1)}}$.

We searched for multiplet candidates and we only found a few examples of split dipole modes in {\it Gaia} DR2\,5233348350879239680. We marked these peaks with '*' in Figure\,\ref{fig:modeids}. In Table\,\ref{tab:table9} we added an azimuthal order assignment according to the splitting. If a single splitting is measured, f$_2$, f$_3$ and f$_4$, f$_5$ we arbitrarily chose one of the peaks to be the central ({\it m}\,=\,0) component. In case of f$_{12}$, f$_{13}$, the splitting is doubled so we may have detected only the side components. Assuming our multiplet identification is correct the average rotation period equals 5.238(48)\,days. The non-split modes are arbitrarily chosen to central components {\it m}\,=\,0.

The multiplet detection helps to constrain the modal degree and provides a head start for determining the asymptotic period spacing, as these three pairs of peaks were assigned {\it l}\,=\,1 modes. In all other cases, including four other stars, we found no additional hints of a modal degree preference and we decided to assign {\it l}\,=\,1 to any peaks that are spaced by a period spacing between 200 and 300\,sec. Those peaks not satisfying our requirements were assigned either {\it l}\,=\,2 or a trapped {\it l}\,=\,1 mode. The latter happened only in {\it Gaia} DR2\,5391470732981138816. The average period spacings of dipole modes we measured in these stars are 288.20\,(1.08)\,s ({\it Gaia} DR2\,5233348350879239680), 279.50\,(35)\,s ({\it Gaia} DR2\,5250674622612902912), 274.30\,(1.26)\,s ({\it Gaia} DR2\,5365175740610237568), 267.84\,(78)\,s ({\it Gaia} DR2\,5391470732981138816), 248.74\,(1.29)\,s ({\it Gaia} DR2\,6096794282417069696). We included the resultant radial order and modal degree assignment in Tables\,\ref{tab:table9} -- \ref{tab:table13}.

\begin{table}
\centering
\caption{List of frequencies detected in {\it Gaia}\,DR2\,5233348350879239680.}
\label{tab:table9}
\resizebox{\columnwidth}{!}{\begin{tabular}{cccccccc}
\hline
\multirow{2}{*}{ID} & Frequency & Period & Amplitude & \multirow{2}{*}{S/N} & \multirow{2}{*}{\it l} & \multirow{2}{*}{\it m} & \multirow{2}{*}{n}\\
&[$\upmu$Hz] & [s] & [ppt] &&\\
\hline
f$_{\rm 1}$ & 76.667(8) & 13043.4(1.4) & 0.824(57) & 12.5 & 1 & 0 & 46\\
\cline{6-8}
f$_{\rm 2}$ & 85.977(11) & 11631.1(1.5) & 0.575(57) & 8.7 & 1 & 0 & 41\\
f$_{\rm 3}$ & 87.073(9) & 11484.6(1.2) & 0.716(57) & 10.8 & 1 & -1 & 41\\
\cline{6-8}
f$_{\rm 4}$ & 90.534(6) & 11045.5(7) & 1.143(57) & 17.3 & 1 & 0 & 39\\
f$_{\rm 5}$ & 91.653(12) & 10910.7(1.5) & 0.529(57) & 8.0 & 1 & -1 & 39\\
\cline{6-8}
f$_{\rm 6}$ & 99.142(11) & 10086.5(1.1) & 0.585(57) & 8.9 & 2 & 0 & 62\\
f$_{\rm 7}$ & 101.635(11) & 9839.1(1.1) & 0.572(57) & 8.7 & 1 & 0 & 35\\
f$_{\rm 8}$ & 104.81(2) & 9541.2(1.8) & 0.331(57) & 5.0 & 1 & 0 & 34\\
f$_{\rm 9}$ & 107.499(9) & 9302.4(8) & 0.714(57) & 10.8 & 1 & 0 & 33\\
f$_{\rm 10}$ & 141.584(3) & 7063.0(2) & 2.096(57) & 31.7 & 1 & 0 & 25\\
f$_{\rm 11}$ & 142.361(19) & 7024.4(9) & 0.35(57) & 5.3 & - & - & -\\
\cline{6-8}
f$_{\rm 12}$ & 147.615(9) & 6774.40(42) & 0.718(57) & 10.9 & 1 & +1 & 24\\
f$_{\rm 13}$ & 149.814(9) & 6674.94(40) & 0.728(57) & 11.0 & 1 & -1 & 24\\
\cline{6-8}
f$_{\rm 14}$ & 171.236(3) & 5839.9(1) & 2.522(57) & 38.2 & 1 & 0 & 21\\
f$_{\rm 15}$ & 177.34(2) & 5638.8(6) & 0.324(57) & 4.9 & 2 & 0 & 35\\
f$_{\rm 16}$ & 183.17(1) & 5459.57(30) & 0.646(57) & 9.8 & 2 & 0 & 34\\
f$_{\rm 17}$ & 190.303(13) & 5254.79(35) & 0.518(57) & 7.8 & 1 & 0 & 19\\
f$_{\rm 18}$ & 219.98(2) & 4545.97(41) & 0.334(57) & 5.1 & - & - & -\\
f$_{\rm 19}$ & 228.411(25) & 4378.08(48) & 0.261(57) & 4.0 & 1 & 0 & 16\\
\hline
\end{tabular}}
\end{table}

\begin{table}
\centering
\caption{List of frequencies detected in {\it Gaia}\,DR2\,5250674622612902912.}
\label{tab:table10}
\begin{tabular}{ccccccc}
\hline
\multirow{2}{*}{ID} & Frequency & Period & Amplitude & \multirow{2}{*}{S/N} & \multirow{2}{*}{\it l} & \multirow{2}{*}{n}\\
&[$\upmu$Hz] & [s] & [ppt] &&\\
\hline
f$_{\rm 1}$ & 41.327(16) & 24197(10) & 0.224(47) & 4.7 & 1 & 87\\
f$_{\rm 2}$ & 47.94(11) & 20859.4(4.9) & 0.325(47) & 6.9 & 1 & 75\\
f$_{\rm 3}$ & 64.508(15) & 15501.9(3.7) & 0.24(5) & 5.1 & 1 & 56\\
f$_{\rm 4}$ & 82.56(1) & 12112.1(1.4) & 0.384(47) & 8.2 & 2 & 75\\
f$_{\rm 5}$ & 103.236(5) & 9686.5(5) & 0.706(47) & 15.0 & 1 & 35\\
f$_{\rm 6}$ & 109.111(12) & 9165.0(1.0) & 0.306(47) & 6.4 & 2 & 57\\
f$_{\rm 7}$ & 125.473(3) & 7969.9(2) & 1.28(5) & 27.1 & 1 & 29\\
f$_{\rm 8}$ & 145.12(1) & 6890.8(5) & 0.36(5) & 7.6 & 1 & 25\\
f$_{\rm 9}$ & 152.038(17) & 6577.3(7) & 0.215(47) & 4.6 & 1 & 24\\
f$_{\rm 10}$ & 166.346(13) & 6011.6(5) & 0.28(5) & 5.9 & - & -\\
f$_{\rm 11}$ & 166.56(1) & 6003.7(4) & 0.356(47) & 7.5 & - & -\\
f$_{\rm 12}$ & 166.766(2) & 5996.4(1) & 1.913(47) & 40.6 & 1 & 22\\
f$_{\rm 13}$ & 166.982(8) & 5988.6(5) & 0.466(47) & 9.9 & - & -\\
f$_{\rm 14}$ & 214.713(13) & 4657.4(3) & 0.278(47) & 5.9 & 1 & 17\\
\hline
\end{tabular}
\end{table}

\begin{table}
\centering
\caption{List of frequencies detected in {\it Gaia}\,DR2\,5365175740610237568.}
\label{tab:table11}
\begin{tabular}{ccccccc}
\hline
\multirow{2}{*}{ID} & Frequency & Period & Amplitude & \multirow{2}{*}{S/N} & \multirow{2}{*}{\it l} & \multirow{2}{*}{n}\\
&[$\upmu$Hz] & [s] & [ppt] &&\\
\hline
f$_{\rm 1}$ & 84.074(13) & 11894.3(1.8) & 0.412(42) & 8.4 & 1 & 44\\
f$_{\rm 2}$ & 99.793(18) & 10020.7(1.8) & 0.291(42) & 5.9 & 1 & 37\\
f$_{\rm 3}$ & 101.038(23) & 9897.3(2.3) & 0.228(42) & 4.6 & - & -\\
f$_{\rm 4}$ & 112.816(4) & 8864.02(33) & 1.238(42) & 25.1 & 1 & 33\\
f$_{\rm 5}$ & 128.048(14) & 7809.6(9) & 0.377(42) & 7.7 & 1 & 29\\
f$_{\rm 6}$ & 132.607(15) & 7541.1(9) & 0.349(42) & 7.1 & 1 & 28\\
f$_{\rm 7}$ & 136.978(14) & 7300.4(8) & 0.371(42) & 7.5 & 1 & 27\\
f$_{\rm 8}$ & 143.537(19) & 6966.9(9) & 0.276(42) & 5.6 & 1 & 26\\
f$_{\rm 9}$ & 168.952(3) & 5918.85(9) & 2.053(42) & 41.6 & 1 & 22\\
f$_{\rm 10}$ & 173.128(24) & 5776.1(8) & 0.219(42) & 4.4 & - & -\\
f$_{\rm 11}$ & 175.858(2) & 5686.40(7) & 2.457(42) & 49.8 & 2 & 36\\
f$_{\rm 12}$ & 177.531(29) & 5632.8(9) & 0.181(42) & 3.7 & - & -\\
f$_{\rm 13}$ & 177.668(2) & 5628.48(6) & 2.768(42) & 56.1 & 1 & 21\\
f$_{\rm 14}$ & 193.201(7) & 5175.94(19) & 0.726(42) & 14.7 & 2 & 33\\
f$_{\rm 15}$ & 196.89(1) & 5078.75(26) & 0.527(42) & 10.7 & - & -\\
f$_{\rm 16}$ & 198.388(7) & 5040.64(17) & 0.807(42) & 16.4 & 1 & 19\\
f$_{\rm 17}$ & 202.604(25) & 4935.7(6) & 0.213(42) & 4.3 & - & -\\
f$_{\rm 18}$ & 224.079(22) & 4462.71(43) & 0.243(42) & 4.9 & 1 & 17\\
\hline
\end{tabular}
\end{table}

\begin{table}
\centering
\caption{List of frequencies detected in {\it Gaia}\,DR2\,5391470732981138816.}
\label{tab:table12}
\begin{tabular}{ccccccc}
\hline
\multirow{2}{*}{ID} & Frequency & Period & Amplitude & \multirow{2}{*}{S/N} & \multirow{2}{*}{\it l} & \multirow{2}{*}{n}\\
&[$\upmu$Hz] & [s] & [ppt] &&\\
\hline
f$_{\rm 1}$ & 66.562(19) & 15023.6(4.2) & 0.65(1) & 5.4 & 1 & 57\\
f$_{\rm 2}$ & 81.086(11) & 12332.6(1.6) & 1.12(1) & 9.3 & 1 & 47\\
f$_{\rm 3}$ & 88.754(21) & 11267.1(2.7) & 0.57(1) & 4.7 & 1 & 43\\
f$_{\rm 4}$ & 97.621(15) & 10243.7(1.6) & 0.82(1) & 6.8 & 1 & 39\\
f$_{\rm 5}$ & 103.09(12) & 9700.3(1.2) & 0.99(1) & 8.2 & 1 & 37\\
f$_{\rm 6}$ & 105.592(22) & 9470.5(2.0) & 0.54(1) & 4.5 & - & -\\
f$_{\rm 7}$ & 106.448(19) & 9394.2(1.7) & 0.64(1) & 5.3 & 1 & 36\\
f$_{\rm 8}$ & 121.176(21) & 8252.5(1.4) & 0.59(1) & 4.9 & 2 & 54\\
f$_{\rm 9}$ & 131.49(2) & 7605.4(1.1) & 0.62(1) & 5.1 & 2 & 50\\
f$_{\rm 10}$ & 140.108(18) & 7137.4(9) & 0.67(1) & 5.5 & 2 & 47\\
f$_{\rm 11}$ & 141.623(17) & 7061.0(8) & 0.72(1) & 5.9 & 1 & t\\
f$_{\rm 12}$ & 143.025(22) & 6991.8(1.1) & 0.54(1) & 4.5 & 1 & 27\\
f$_{\rm 13}$ & 169.373(12) & 5904.14(42) & 1.02(1) & 8.4 & 1 & 23\\
\hline
\end{tabular}
\end{table}

\begin{table}
\centering
\caption{List of frequencies detected in {\it Gaia}\,DR2\,6096794282417069696.}
\label{tab:table13}
\begin{tabular}{ccccccc}
\hline
\multirow{2}{*}{ID} & Frequency & Period & Amplitude & \multirow{2}{*}{S/N} & \multirow{2}{*}{\it l} & \multirow{2}{*}{n}\\
&[$\upmu$Hz] & [s] & [ppt] &&\\
\hline
f$_{\rm 1}$ & 98.741(18) & 10127.5(1.8) & 1.582(92) & 15.2 & 1 & 41\\
f$_{\rm 2}$ & 138.863(19) & 7201.3(1.0) & 1.535(92) & 14.7 & 1 & 29\\
f$_{\rm 3}$ & 148.208(47) & 6747.3(2.1) & 0.607(92) & 5.8 & 2 & 46\\
f$_{\rm 4}$ & 161.876(11) & 6177.57(43) & 2.496(92) & 23.9 & 1 & 25\\
f$_{\rm 5}$ & 235.547(24) & 4245.44(43) & 1.201(92) & 11.5 & 2 & 29\\
f$_{\rm 6}$ & 239.563(36) & 4174.3(6) & 0.786(92) & 7.5 & 1 & 17\\
f$_{\rm 7}$ & 272.763(49) & 3666.2(7) & 0.578(92) & 5.5 & 1 & 15\\
\hline
\end{tabular}
\end{table}

\section{Summary}
\label{summary}
To date, the sample of sdBVs has been selected by random discoveries made from ground-based and space observations. This limits the study of sdBVs to individual cases only and it does not allow for inferring how pulsation properties depend on a stellar population. We undertook a search for sdBVs candidates selected from a sdB database \citet{geier20} and based on an all-sky photometry collected during the {\it TESS} mission in the LC mode. This data still limits our search to g-mode sdBVs, which is a consequence of a 30\,min. cadence, however the search utilizes the most updated sdB database along with the only all-sky space survey, allowing for the most complete sample of g-mode sdBVs currently possible.

Many sdBV candidates are allocated for {\it TESS} SC monitoring and these objects have time-series data pulled out and ready for variability check. We focused on additional sample of sdBV candidates that are located on {\it TESS} silicons but with no automated time-series data pulled out. We have prepared and used a set of applications and scripts that allowed us to derive the fluxes and to calculate amplitude spectra to detect a significant flux variation.

From time-series data and amplitude spectra, we detected significant variability in 1807 (out of 20,642) objects. 28 objects are classified as sdBs and only two of them show convincing pulsations. One object shows eclipses and a reflection effect typical of HW\,Vir systems, while the remaining 25 sdBs show significant signal in their amplitude spectra, which we interpreted as a binary signature. A sample of 77 objects not classified as sdBs were found to be variable and we include them in this paper (on-line material) for completeness and as a resource for others. The remaining 1702 objects were found to be variable but are spectroscopically unclassified, hence we are unable to verify if they are sdBVs or another type of variable stars yet. In this group we found 83 objects showing a significant signal typical of g-modes in sdBVs. We selected the five objects best-suited for mode identification and assigned modal degrees. We used multiplets and a period spacing for this purpose. The sequences of presumably same degree overtones are not too complete but the multiples of 250\,sec (ish) can still be found. This may be another argument for these objects being sdB stars. However, our identification will only be reliable if these objects are spectroscopically confirmed to be sdBs. Regardless the correctness of our choice, other objects are not suitable for the mode identification since the signal we detected either has too few peaks or the peaks are too close to the Nyquist frequency. 

We also found 83 eclipsing binaries and organized them into three subgroups based on their eclipse content. We found 32 objects that show both primary and secondary eclipses, 42 objects that show only primary eclipses, and nine objects that show secondary eclipses out of 0.5 phase, indication of eccentric orbits. The last sample of variables contains 1535 objects that show either binary signatures or classical pulsator asymmetric light curves. 273 objects show flux variation amplitudes that are large enough to see in phased time-series data. These objects show one symmetric maximum, one asymmetric maximum, and two maxima. The remaining 1262 objects show amplitude spectra that are consistent with binarity.

We also detected flux variations in two known novae, of which only one is spectroscopically classified.

Our search for variable sdB stars in TESS LC data reveals a few new and more than a thousand candidates for sdBVs. We used our discoveries to propose those candidates to be observed in either 2\,min or 20\,sec cadence, during the second run of TESS in the southern ecliptic hemisphere. If these objects are allocated the upcoming data may bring additional discoveries of p-modes, confirming their sdB nature. Our work is the first focused on an all-sky TESS survey to search for sdBVs and, when the spectral classification is performed, we consider it to be the most updated list of sdBVs in the southern ecliptic hemisphere. Such a sample will be very useful to understand the pulsation--population relationship.

\section*{Acknowledgements}
Financial support from the Polish National Science Center under projects No.\,UMO-2017/26/E/ST9/00703 and UMO-2017/25/B ST9/02218 is acknowledged. This paper includes data collected by the \tess\ mission. Funding for the \tess\ mission is provided by the NASA Explorer Program. This work has made use of data from the European Space Agency (ESA) mission {\it Gaia} (\url{https://www.cosmos.esa.int/gaia}), processed by the {\it Gaia} Data Processing and Analysis Consortium (DPAC, \url{https://www.cosmos.esa.int/web/gaia/dpac/consortium}). Funding for the DPAC has been provided by national institutions, in particular the institutions participating in the {\it Gaia} Multilateral Agreement. Fruitful remarks from an anonymous referee are appreciated.

\section*{Data availability}
The datasets were derived from MAST in the public domain archive.stsci.edu.

%%%%%%%%%%%%%%%%%%%%%%%%%%%%%%%%%%%%%%%%%%%%%%%%%%

%%%%%%%%%%%%%%%%%%%% REFERENCES %%%%%%%%%%%%%%%%%%

% The best way to enter references is to use BibTeX:

\bibliographystyle{mnras}
\bibliography{bibliography} % if your bibtex file is called example.bib

%\printbibliography
% Alternatively you could enter them by hand, like this:
% This method is tedious and prone to error if you have lots of references
%\begin{thebibliography}{99}
%\bibitem[\protect\citepauthoryear{Author}{2012}]{Author2012}
%Author A.~N., 2013, Journal of Improbable Astronomy, 1, 1
%\bibitem[\protect\citepauthoryear{Others}{2013}]{Others2013}
%Others S., 2012, Journal of Interesting Stuff, 17, 198
%\end{thebibliography}

%%%%%%%%%%%%%%%%%%%%%%%%%%%%%%%%%%%%%%%%%%%%%%%%%%

%%%%%%%%%%%%%%%%% APPENDICES %%%%%%%%%%%%%%%%%%%%%

%\appendix

%\section{Some extra material}

%If you want to present additional material which would interrupt the flow of the main paper,
%it can be placed in an Appendix which appears after the list of references.

%%%%%%%%%%%%%%%%%%%%%%%%%%%%%%%%%%%%%%%%%%%%%%%%%%

% Don't change these lines
\bsp	% typesetting comment
\label{lastpage}
\end{document}